\documentclass{svjour3}                     % onecolumn (standard format)
\smartqed  % flush right qed marks, e.g. at end of proof
\usepackage{graphicx}
\begin{document}

\title{Instruments of RT-2 Experiment onboard CORONAS-PHOTON
and their test and evaluation III: Coded Aperture Mask and Fresnel Zone Plates in RT-2/CZT Payload
\thanks{This work was made possible in part from a grant from Indian Space Research Organization 
(ISRO). 
Upendra Desai initiated the Zone Plate work and his contribution in the form of suggestions and 
discussions has proved quite valuable for this work. 
The whole-hearted support from G. Madhavan Nair, Ex-Chairman, ISRO, who initiated the RT-2 project, is 
gratefully acknowledged.
}
}

%\subtitle{Zone Plates for X-ray Imaging\\ If so, write it here}

\titlerunning{Fresnel Zone Plates for RT-2}

\author{Anuj Nandi \and S. Palit \and D. Debnath \and Sandip K. Chakrabarti \and T. B. Kotoch \and 
R. Sarkar \and Vipin K. Yadav \and V. Girish \and A. R. Rao \and D. Bhattacharya}

\authorrunning{Nandi et al. } % if too long for running head

\institute{  A. Nandi$^+$, S. Palit, D. Debnath, T. B. Kotoch, R. Sarkar and Vipin K. Yadav$^+$ \at
Indian Centre for Space Physics, 43 Chalantika, Garia Station Road, Kolkata 700084\\
Tel.: +91-33-24366003\\
Fax: +91-33-24622153 Ext. 28\\
\email{anuj@csp.res.in; sourav@csp.res.in; dipak@csp.res.in; tilak@csp.res.in; ritabrata@csp.res.in;
vipin@csp.res.in}   
($+$: Posted at ICSP by Space Science Division, ISRO Head Quarters)
\and
Sandip K. Chakrabarti \at
S.N. Bose National Centre for Basic Sciences, JD Block, Salt Lake, Kolkata 700097 \\
(Also at Indian Centre for Space Physics, 43 Chalantika, Garia Station Rd., Kolkata 700084)\\
Tel.: +91-33-23355706\\
Fax: +91-33-23353477\\
\email{chakraba@bose.res.in}           %  \\
%   \emph{Present address:} of F. Author  %  if needed
\and
V. Girish \at
Space Astronomy and Instrumentation Division, ISAC, Bangalore, 560017 \\
\email{giri@isac.gov.in}           %  \\
\and
A. R. Rao \at
Tata Institute of Fundamental Research, Homi Bhabha Road, Colaba, Mumbai, 400005\\
              \email{arrao@tifr.res.in}   
\and
D. Bhattacharya \at
Inter-University Centre for Astronomy and Astrophysics, Ganeshkhind, Pune, 411007\\
	\email{dipankar@iucaa.ernet.in}
}
\date{Received: date / Accepted: date}
% The correct dates will be entered by the editor
\maketitle

\begin{abstract}

Imaging in hard X-rays of any astrophysical source with high angular resolution is a
challenging job. Shadow-casting technique is one of the most viable options for imaging in hard
X-rays. We have used two different types of shadow-casters, namely, Coded
Aperture Mask (CAM) and Fresnel Zone Plate (FZP) pair and two types
of pixellated solid-state detectors, namely, CZT and CMOS in RT-2/CZT payload, 
the hard X-ray imaging instrument onboard the
CORONAS-PHOTON satellite. In this paper, we 
present the results of simulations with different combinations of coders (CAM \& FZP) and 
detectors that are employed in the RT-2/CZT payload.
We discuss the possibility of detecting transient Solar flares with 
good angular resolution for various combinations. Simulated results are compared with 
laboratory experiments to verify the consistency of the designed configuration.

% \subclass{MSC code1 \and MSC code2 \and more}
\end{abstract}

\section{Introduction}
\label{intro}

%Magnetic reconnection on the surface of Sun impulsively releases energy causing solar flares.
% A detail study of solar flare continium spectrum and imaging of them has the potential of gaining 
%new insights on the physical processes as well as the emission mechanisms.
% During solar flares electrons are accelerated upto hundreds of MeV. Bremstrahlung emission from 
%non thermal electrons produces the radiation upto 10s of MeV. It has been seen that non thermal 
%X-ray spectra in flares often have a relatively sharp downword break in the range  30 -50 keV for 
%small flares but as high as greater than 100 keV in large flares.

%So study of the hard X-ray emanating during the occurence of solar flares is essential.

Imaging astrophysical sources in X-rays of electro-magnetic wave band, 
specially in hard X-rays, 
is really a challenging and difficult task. X-ray photons do not efficiently
get reflected or refracted because of their ability to penetrate deep into the interacting material.
It is, however, possible to image with soft X-ray photons by making them hit a highly polished mirror
at a small angle to the reflecting surface. This process, called grazing incidence imaging technique, 
is effective mainly for soft X-rays (up to $\sim$ 10 keV). 
For example, in {\it Chandra} and {\it XMM-Newton} missions, soft X-ray imaging has been done very 
successfully with this technique. Unfortunately, this direct imaging technique is
very difficult to implement for 
hard X-rays and $\gamma$-rays.

Instead of direct imaging, some indirect imaging techniques have been developed and adopted for 
imaging in hard X-rays as well as in $\gamma$-rays. Shadow casting method is one such efficient 
indirect imaging technique (Mertz \& Young 1961). This technique is based on the total absorption of 
hard X-rays by the shadow-caster, allowing the generation of patterns in 
the detector plane, which can be deconvolved to get the source image.
There are various types of shadow casting methods depending on the structure of coder and mask
pattern. One such approach is to use a single plane coder, namely Coded Aperture Mask (CAM),
which is widely used in different astronomical observations (Dicke 1968, Ables 1968, Baker et al., 
1983 and Caroli et al., 1987). In this approach, mask pattern is basically followed by the individual 
pixel dimension of the detector and therefore, the angular resolution depends on the separation between 
the coder and the detector plane and the smallest size of the coder element.

Another approach (Mertz \& Young 1961, Desai et al., 1998, Chakrabarti et al., 2009) to image in hard 
X-rays with high angular resolution is to use double plane coder of Fresnel Zone Plates (FZPs). 
The high angular resolution, which is achievable upto a few arc-sec through this method depends 
on the outermost zone width and separation between the two coders (pair of FZP) (Palit et al., 2009). 

RT-2/CZT payload, one of the most important instruments of the RT-2 experiment (Debnath et al. 2009, 
Kotoch et al. 2009, Sarkar et al. 2009, Sreekumar et al. 2009) onboard 
CORONAS-PHOTON mission (Kotov et al. 2008, Nandi et al. 2009), is specially designed to 
provide high 
resolution spectral and imaging information of hard X-ray solar flares at energies above 20 keV.
For this, we have employed both the shadow casting methods -
the Coded aperture mask (CAM) and dual Fresnel zone plate (FZP) as a coder in RT-2/CZT payload. 
Detection of hard X-rays requires interaction of photons with denser and 
heavy material and for imaging, position sensitivity is also essential for the detector. 
We have used two types of solid-state pixellated detectors, namely, Cadmium Zinc Telluride (CZT) 
and Complementary Metal Oxide Semiconductor (CMOS) detectors in RT-2/CZT payload. The solid-state 
CZT detector has a very good combination of energy resolution, detection efficiency and
it is  best suited 
with CAM pattern in terms of required spatial resolution. On the other hand, the CMOS detector, with 
high positional accuracy, is very good for imaging purposes and with the combination of FZP it can 
give a few arc-sec of angular resolution. 

X-ray observation of solar flares must be made from a platform launched above the atmosphere. 
Sending such instrument to near space by balloon is an option. But the best way is to send it 
through a satellite orbiting the Earth. Many missions, like {\it GRANAT}, {\it HETE}, {\it SWIFT} etc. 
had included shadow casting method for imaging in hard X-rays.

%Sigma, XRT, HXT, ART-P, COMIS -TTM etc are such instrumentsn which are of no more use now. 
%XTE-ASM, HETE-WXM, SWIFT-BAT are some example of some space based instrument still operating 
%and  employing shadow casting imaging. 
All of these mission employ CAM 
and they are for imaging mainly of celestial objects 
like AGNs, compact objects and GRBs. RT-2 will be the first of its kind to use both 
types of shadow caster (CAM \& FZP) for studying hard X-ray solar flares.
 
In this paper, we discuss in detail the coders and detectors employed in the 
RT-2/CZT payload. We also provide simulation results along with some laboratory test 
results to validate these simulations.
In the next section, we discuss the principle of operation of CAM and FZP as shadow casters and their basic 
constructions. This includes the theory behind the working principle and the image reconstruction 
processes. In the same section the specifications of the instruments, viz, detectors and 
shadow casters are given. Then in section 3, we discuss the angular resolution (AR) and field of 
view (FOV) of all four combinations of coder and detector used in the payload and make comparative 
study of them in terms of these two important characteristics.
In section 4, we present the simulation aspects of each configuration used in this payload. 
These include shadow formations (eg. Moir\'{e} fringe for FZPs) and source reconstruction for one or 
more sources. In section 5, we discuss what we expect in context of solar observation from the 
instrument used in this experiment. The results of the laboratory experiments carried out with the 
designed instruments 
are presented in section 6. Finally, in section 7, we make some concluding remarks.  

\section{CAM and FZP as shadow masks}

Coded mask imaging is a class of spatial multiplexing technique (Caroli et al., 1987) for imaging 
of objects in high energy part of the electromagnetic spectrum, specially in X-rays \& $\gamma$-rays. 
It is a two step process in which 
data is acquired from some part of the sky (within FOV of the instrument) and source image is
reconstructed by some computation procedures.
A coded mask is a plate consisting of areas which are transparent and opaque to 
photons within certain energy ranges. The transparent and opaque areas or mask elements are 
generally of 
equal size and are distributed in a predetermined pattern. The mask acts as a shadow caster, viz, 
shadows of the mask are projected on the detector placed below the mask facing the sky, 
by the rays coming from the sources in the visible part of the sky. The projected shadow on
the detector plane has the same coding information
as that of the mask pattern. The information on the directions of the rays falling on the mask and 
hence the source positions in the field of view are encoded in the amounts of shift of all the shadow 
patterns with respect to the central position and the information on the source intensities are 
encoded in the strengths of the patterns. 

%Among different types of coded mask proposed, only few, viz, Coded Aperture Mask (CAM) (Dicke 1968, 
%Ables 1968, Fenimore et al., 1979) and Fresnel Zone Plate (FZP) coder (Mertz 1965) satisfy the 
%conditions of proper performance in indirect imaging of sky. The conditions are that the 
%autocorrelation function of the mask pattern should be a delta function, i.e, 
%having a single peak with flat side lobes and there should be an optimum signal to noise ratio. 

\subsection{Coded Aperture Mask (CAM)}

\subsubsection{CAM as a coder}

A pinhole camera has the characteristics required for a proper indirect imaging performance, but 
it has poor signal-to-noise ratio.
The sensitivity (signal to noise - S/N -  ratio) may be increased by increasing the pinhole area but at the same time it will
degrade the angular resolution of the device.
A random pinhole camera (Dicke 1968 and Ables 1968) can be constructed by placing many 
pinholes at random in a plate. 
This increases the open area of the plate, required for better sensitivity and also preserves the 
angular resolution. It is found that ideal patterns for random pinhole camera should be based on 
cyclic difference sets (Gunson \& Polychronopulos 1976, Fenimore \& Cannon 1978). It is also required 
that every detector pixel is exposed to one full mask pattern.
These patterns are also called Uniformly redundant arrays (URA) (Fenimore \& Cannon 1978).

In RT-2/CZT, CZT detector modules of 256 pixels of size $\sim$2.5 $\times$ 2.5 mm$^2$ 
are used (see Kotoch et al. 2009 for details). For good hard X-ray spectroscopic
observations, it is necessary to have individual pixel calibration and simultaneous
background measurements. Hence, the individual pixel reading capability of
these modules can be effectively used by incorporating an appropriate
coder for the purpose of simultaneous background measurement. Further,
solar hard X-ray emission (above 20 keV) is observable only during
hard X-ray flares which generally are associated with some active regions.
Hence an identification
of the active region on the solar surface during solar hard X-ray flares,
correct to arc-minute accuracy, would also be a desirable characteristics of the
coder. 
%URA can be constructed for a level of transmission upto 50 $\%$. 
A pseudo noise Hadamard set
which is capable of giving maximum possible transmission (50 $\%$) (Caroli et al., 1987, In't Zand 
1992) is suitable for a mask employing CZT detector module. For the detector of 
256 pixels, we chose 256 mask elements (to maximize the localization accuracy). Since
a pseudo noise Hadamard set has 2$^m$-1 elements, we chose m = 8 with an extra 
opaque element in the mask.

A pseudo
noise Hadamard set is constructed from a shift-register algorithm (Peterson 1961). 
The coefficients $p_j$ of a m$^{th}$ order primitive polynomial (all the coefficients are either 1 or 0) 
can be used as the generating function of a mask pattern of length 2$^m$-1, provided the polynomial is also an irreducible one.
%he coefficients  $p_j$ of an m$^{th}$ degree polynomial can be chosen 
%rbitrarily and the
The mask elements $a_i$ where i = 0, 2$^m$-2 can be generated from the coefficients by using the shift register algorithm 
%
%Let us have a polynomial in $x$ as follows (Bhattacharya et al. ****),
%$$
%P = \sum_{j = 0}^m p_{j} \times x^{j},
%\eqno{(1)}
%$$
%which represents a polynomial of degree $m$ in $x$. From these, primitive polynomials with coefficients
%($p_{j}$s) having values 0 and 1 can be chosen. Among these primitive polynomials irreducible ones 
%having degree $m$ can be used as generating function of a mask pattern of length $n = 2m - 1$.
%Let this be represented by a sequence,
%$$
%a_{i}, i = 0,.....,(2^{m}-2),
%$$
%where $a_{i}$s are either 0 or 1, with 1-s and 0-s corresponding to open and closed mask elements respectively.
%The first $m$ elements can be chosen arbitrarily. The following elements are then 
%generated using shift register 
%algorithm
$$
a_{i+m} = \sum_{j = 0}^{m-1}p_{j} \times{a_{i+j}}(mod~2).
\eqno{(1)}
$$
The above mentioned process for m = 8 gives 16 polynomials from which CAM patterns may be generated. 
Out of these 16 options, the following 
two polynomials were chosen for the CZT-CAM configurations:
$$
                         x^8 +x^6 +x^5 + x^2 + 1
$$
and
$$
                         x^8 +x^4 +x^3 + x^2 + 1.
$$
The choice was made keeping in mind that each individual pixel should get possible maximum 
amount of mechanical support. These were determined by minimizing the mechanical support parameter (MSP) defined as,
%
%comparing the figure of merit (FOM) of each 
%of the sixteen patterns. The final FOM of the patterns are found by the relation 
$$
MSP = b + 4c + 8d,
$$
where $b$ corresponds to the number of segments of each pattern held at two or more corners,
$c$ and $d$ corresponds to the number held respectively at `one' and `no' corner.
The `one' and `no' corner elements have to be mechanically held to the
neighboring closed elements, slightly reducing the open element area
and hence reducing the sensitivity.  The 
two patterns generated by the above two polynomials have the minimum MSP. For the 1$^{st}$ one 
$$
 b = 3,~ c = 2,~ d = 0,
$$
and for the 2$^{nd}$ one
$$
b = 7,~ c = 1,~ d = 0.
$$
And each of which gives MSP = 11.
%
%Two patterns generated by the two polynomials given above correspond to minimum figure 
%of merit. In such URA, the plate area is equally and pseudo-randomly divided into opaque and 
%transparent regions (squares). 
The two CAM patterns that are used in this payload are shown in Figure 
4(a) and Figure 6(a). 
%These patterns obey the most important condition regarding autocorrelation 
%function. An advantageous fact about CAM is that it is a single plate coder and can be easily 
%constructed.

\subsubsection{Image reconstruction}

The transformation from object distribution function in the sky, S(x,y) to the
spatial distribution of the flux in the detector plane, D(x,y) can be mathematically written as 
(Caroli et al., 1987),
$$
D(x,y) = C(x,y) \ast S(x,y) + B(x,y),
\eqno{(2)}
$$
where $C(x,y)$ is the aperture transmission function, $B(x,y)$ is signal independent noise term  
and $\ast$ is the convolution operator. An estimation of the sky function ($S'(x,y)$) can then be 
made by filtering the detector flux distribution by a suitable decoding function $K(x,y)$ such that 
$K(x,y)*C(x,y)$ is a delta function. Therefore, the sky function can be written as,
$$
S'(x, y) = K(x,y) \ast C(x,y)  \ast S(x,y) + K(x,y) \ast B(x,y),
\eqno{(3)}
$$
which is the reconstructed source distribution in the detector plane.

There are different types of reconstruction codes for CAMs, out of which suitable ones are chosen 
fulfilling the requirements of source informations.
%In our reconstruction processes, we consider $C(x,y)$ as a matrix containing elements 1 and 0 and 
%dimension governed by the number of elements of detector function and opaque and transparent 
%CAM element distribution array taken as matrices.
In our reconstruction process, all the above mentioned functions are represented as matrices.
S is taken as a column matrix with number of elements equal 
to the total number of division i.e how the observed part of the sky is divided, and the value of
each element is the source strength of corresponding divisions of the sky. D is the row
matrix with number of elements equal to the number of detector pixels and each element 
representing the counts obtained in the corresponding pixels. Then C is defined as a matrix 
(a similar approach is followed by Caroli et al. 1987) whose number of column is equal to the 
number of elements in S and number of rows is equal to the number of elements of D, and C(i,j) is
equal to 1 or 0 if the line joining the sky pixel corresponding to the j-th element
of S and detector pixel corresponding to i-th element of D passes respectively
through a transparent and opaque region of the CAM.
Thus Eqn. 2 can be written as,

$$
D = C \ast S + B,
%\eqno{(2)}
$$
where, B is a matrix which carries the noise information of the detector. 
Similarly, the sky function can be written in the matrix representation as,

$$
S' = K \ast C \ast S + K \ast B,
%\eqno{(3)}
$$

Here, we use $K$ as the matrix inverse to the matrix $C$. The operator $\ast$ simply denotes the 
matrix product. The distribution $S'$ with some filtering by point spread function (PSF) gives near 
exact reconstruction of the object distribution. Object or source planes reconstructed by the above 
mentioned method from detector flux distributions, is verified with simulations and experimental 
results for various source distributions that are presented in $\S$4 and $\S$6 respectively.  
This method is used in this paper to get an idea of the source localization ability.
For a good estimate of the source intensity, however, we plan to use a shadow fitting 
procedure, which can be fine-tuned to include experimental systematics
like unequal pixel dimensions etc. (Ajello et al., 2008).
%{\bf This reference should be about Swift/BAT deconvolution method}

\subsection{ Fresnel Zone Plate (FZP) pair}

\subsubsection{Dual FZP as a coder}

A Fresnel Zone Plate (FZP) coder is a dual plate coder, where two zone plates are placed at some 
distance apart. Fresnel zone plate has the following transmittance function (Barrett \& Myers 2004, 
Chakrabarti et al., 2009)
$$                                      
T(r) =  {1 \pm  sgn[sin(\alpha r^ 2)]}S (r),
\eqno{(4)}
$$                              
where $\alpha$ is a parameter of the zone plate.
S(r) is a support function (equals to 1 within the outer boundary of the zone plate and 
zero outside). The {\it sgn} function is +1 when sin($\alpha r^ 2$) $>$ 0 and - 1 when 
sin($\alpha r^ 2$) $<$ 0. 
Then the transmission function of the combined zone plates resembles a part of Fourier Transform 
(Mertz 1965) from source plane to detector plane.
Zone plates are constructed such that $\alpha$ is related with n$^{th}$
zone radius $r_{n}$ by any one of the following equations,
$$
\alpha r_{n}^2 = n\pi,
\eqno{(5)}
$$
or
$$
\alpha r_{n}^2 = (n \pm  \frac{1}{2} )\pi.           
\eqno{(6)}
$$ 
Taking two zone plates each of which with $\alpha$ satisfying either Eqn. 5 or Eqn. 6 and with 
+ or - sign in Eqn. 6 (positive and negative zone plate), we can produce transmission function 
resembling any of the four parts of Fourier transform from source plane to detector plane 
(Chakrabarti et al., 2009). 

One of the pair used in this payload (CONFIG-3) is a positive cosine pair, which implies 
that the shadow pattern produced in the detector is the positive cosine part of the Fourier 
transform (Chakrabarti et al., 2009). In this case, the radii of the n$^{th}$ zone of both the 
zone plates are given by 
$$
r_{n}= \sqrt(n)r_{1},
\eqno{(7)}
$$
where $ r_{1}$ is the innermost zone radius and the central zones of the two zone plates are 
transparent.

The other pair (used in CONFIG-4) is a negative cosine pair. For this pair, the radius of 
the n$^{th}$ zone of both the zone plates are given by the same equation as that of 
the previous one 
but both the zone plates are negative in nature, i.e, the innermost zones are opaque to X-rays. 

X-rays passing through two zone plates of a coder produces a shadow in
the detector, called Moir\'{e} pattern. The spacing (S) between two
adjacent fringes in the Moir\'{e} pattern is determined by the
orientation of the rays, hence on the source position in the field of
view and is governed by the relation (Desai et al., 1998),
$$
S = \frac{r_{1}^2}{D~tan\theta},
\eqno{(8)}
$$
where $r_{1}$ is the inner zone radius, $D$ is the spacing between two zone plates and $\theta$
is the off axis angle of the source, i.e, the angle made by the source position with the common 
central axis of the two zone plates.
We can find the source position by measuring the fringe separation in the pattern and
also the information on intensity can be obtained from the strength of the shadow.

In Figure 1(a-b), we present two Moir\'{e} fringe patterns generated by simulations for one of our 
FZP coder configuration (CONFIG-4: FZP2 + CMOS), one of which is for an on-axis source (Figure 1a) 
and the other for an off-axis source (Figure 1b). For 
on-axis source at infinite distance, the rays fall on the front zone plate exactly face on and  
two plates are exactly parallel to each other. So the rear zone plate is exactly shadowed by the 
front one and the fringe pattern resembles exactly a single FZP pattern (Figure 1a). But in 
Figure 1b, as the source is at an off-axis position by an offset of 300 arcsec, straight line 
fringes appear in the pattern. 

\begin{figure}[h]
\centering
\includegraphics[height=2.3in,width=2.3in]{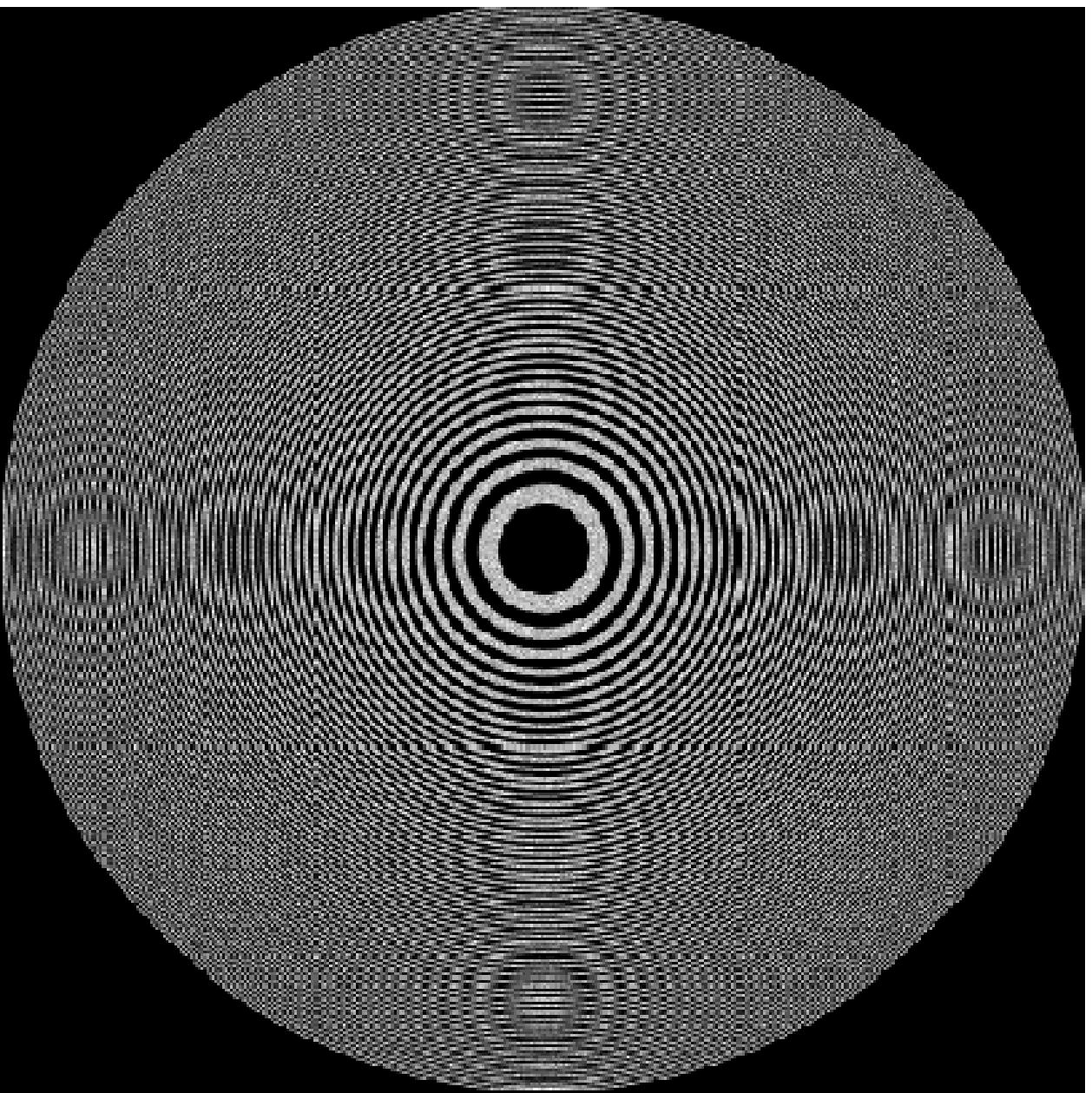}\hspace{0.2 cm}
\includegraphics[height=2.3in,width=2.3in]{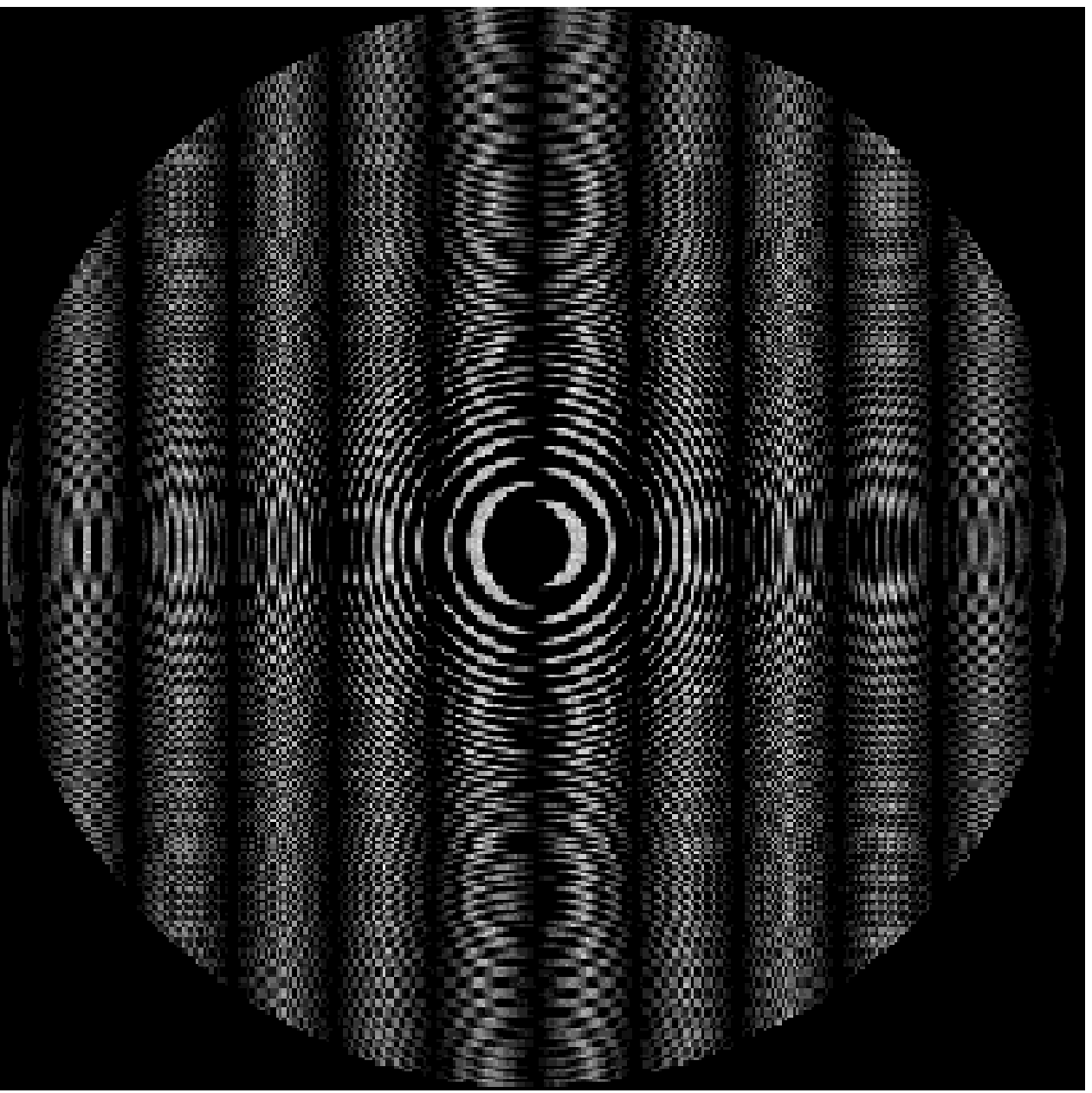}
\caption{Moir\'{e} patterns obtained (simulated) with a cosine negative zone plate pair for an on-axis 
(a) and an off-axis (b) source position.}
\end{figure}

\subsubsection{Image reconstruction}

As the shadow pattern is a part of Fourier transform, we can reconstruct the observed 
part of the sky by applying Inverse Fourier transform on the photon distribution array obtained 
in the detector plane (Mertz 1965). During the image reconstruction, a computer developed FFT code is 
used to do the inverse Fourier transform.

As a single set of FZP coder gives a part of the Fourier transform, the inverse Fourier transform
returns extra objects other than the required source (Chakrabarti et al., 2009). These include a
pseudo source (ghost image) exactly at the position of mirror reflection of the actual source, and a 
central DC-offset. If we use two pairs of zone plates with suitable specifications, we can remove 
the pseudo-source. By using four suitably combined pairs of zone plates, we can also remove the central 
offset (Palit et al., 2009).

For a point source at finite distance, the reconstructed source looks like a dark circular 
spot (as can be seen in Figure 15(b) and Figure 16(b)). This spreading, which 
deteriorates the angular 
resolution of the instrument is due to the divergence of the photon beam incident on the front 
FZP plate (Palit et al., 2009). Diverging effect can be rectified by modifying the zone radii of the 
second FZP plate (Palit et al., 2009).

\subsection{Configuration details}

It is already mentioned that RT-2/CZT payload consists of different types of coders
and detectors for
imaging in hard X-rays of solar flares.
% with achievable spatial angular resolutions. 
In this section,
we present the configuration details of the RT-2/CZT payload. 
The specifications of the detectors used in the payload are given in Table 1.

\subsubsection{Detector specifications}

\begin{center}
\centerline {Table 1}
\begin{tabular}{  c c c  } \hline
{\bf Detectors}  &  CZT (3 numbers) &  CMOS (1 number) \\ 
\hline 
{\bf Dimension (cm)}  & 4 $\times$ 4  & 2.5 $\times$ 2.5 \\
{\bf Number of pixels}   & 16 $\times$ 16  & 512 $\times$ 512   \\
{\bf Pixel dimension}  & (2.5 $\times$ 2.5) $mm^2$ &  (50 $\times$ 50) $\mu^2$  \\ 
{\bf Geometric area}  &  (4 x 4) $cm^2$ $\times$3  &  (2.4 $\times$ 2.4) $cm^2$  \\
\hline 
\end{tabular}
\end{center}
 
\subsubsection{Coder specifications}

We have two different types of shadow-casting devices (coders), namely, CAMs and FZPs. Two different
types of CAM pattern are used for two CZT modules, whereas two pairs of zone plates with different
dimensions are used for one CZT module and CMOS detector. The specifications of the coders those
are used in RT-2/CZT payload are given in the Table 2.

\begin{center}
\centerline {Table 2}
\begin{tabular}{  c c c } \hline
{\bf Device (coder)} &  CAM & Dual FZP  \\ 
\hline 
{\bf Material}  &  Tantalum &  Tungsten  \\    
{\bf Thickness(mm)}  &  0.5  &  1 (each FZP)  \\    

{\bf No. of device}  &  2  &  2  \\    

{\bf No. of plates in a device} & 1 & 2 \\
   
{\bf Area of coder plates} & 16 $cm^2$ & 7.06  $cm^2$ for FZP1 \\
                           &           & 4.52  $cm^2$ for FZP2 \\
  
{\bf Spacing between plates} & ------ &  32 cm for FZP1 \& FZP2 \\
    
{\bf Spacing between } &    40 cm   &  8 cm  \\
{\bf coder and detector} &                 & (lower FZP to detector)  \\
{\bf Coder shape}  &  Square & Circular zones  \\
    
{\bf No. of coder element} & 16 $\times$ 16 = 256 & 151 for FZP1 \\
                           &                        &  144 for FZP2 \\

{\bf Smallest coder size/width} & 2.5 x 2.5 $mm^2$ &  50 $\mu$ for FZP1 \\
                                &                  &  41 $\mu$ for FZP2 \\
\hline
\end{tabular}
\end{center}

FZP1 is a `positive' cosine type and FZP2 is `negative' cosine type coder.
The front view of collimator containing all the shadow-casters (coders) is shown in Figure 2.

\begin{figure}[h]
\centering
\includegraphics[height=3.5in,width=3.5in,angle=90]{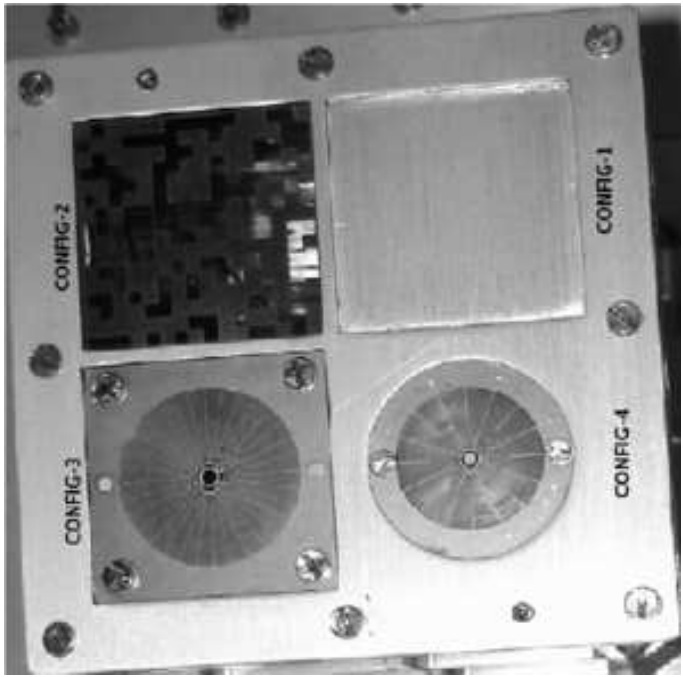}
\caption{Front view of the collimator displaying all the coders. One CAM is shielded with 1 mm Al 
sheet (CONFIG-1) and the other one is open to the sky (CONFIG-2). Front plates of FZP1 (CONFIG-3) and 
FZP2 (CONFIG-4) are also shown.}
\end{figure}

\subsubsection{Detector-Coder combination}

Detectors and coders are placed at the two ends of the collimator of the payload. The collimator
is divided into four quadrants with a height of 32 $cm$. Each quadrant with coder and detector is
an independent configuration and it is termed as {\it CONFIG}. Therefore, RT-2/CZT payload has four
configurations that are summarized in the Table 3.

\begin{center}
\centerline {Table 3}
\begin{tabular}{  c c c c  } \hline
Configuration &  Combination  &  Angular resolution & FOV   \\ \hline
CONFIG-1  & CAM1 + CZT1  & 21.5 arc-minute  & 5.72 degree   \\
CONFIG-2  &  CAM2 + CZT2  &  21.5 arc-minute & 5.72 degree  \\
CONFIG-3  &  FZP1 + CZT3  &  64 arc-sec &  409 arcsec  \\
CONFIG-4  &  FZP2 + CMOS  &  54 arc-sec  &  4.29 degree   \\
\hline
\end{tabular}
\end{center}

\section{Angular resolutions (ARs) and field of views (FOVs) of the coders}

In imaging, angular resolution (AR) and FOV are the crucial aspects 
which determine the sensitivity and effectiveness of the instrument.
%to measure the
%source location more accurately within the FOV (eg. Solar flares in the surface of Sun). 
Generally,
angular resolution of an instrument depends on the smallest coder element size and separation 
between the coder and the detector. In the RT-2/CZT payload we are using various types 
of coders and detectors in different configurations and  we present 
the calculation of the AR and FOV of 
all four configurations. The angular 
resolution of any coder-detector system can be calculated using the schematic diagram 
shown in Figure 3. Let us consider the case
of CAM, where AB is the size of a single CAM element and CD is the size of a detector pixel. Now, 
if two sources have to be resolved by the imaging system, the rays coming from the two sources to 
any pixel must pass through different CAM elements. 
Hence the separation between the two points where
the rays from the two sources (which falls on the same detector pixel) fall on two different CAM 
pixels must not be less than AB.
The angle subtended by this minimum distance on any point of the detector pixel is $\theta_p$.

In case of FZP coder, the scenario would be a little different. Here, AB and CD  would be the radii 
of finest zones of the front and rear zone plates. But as every point of the finest zone is 
equivalent for imaging purpose, the angular resolution should be $\theta_r$ rather than $\theta_p$
(Palit et al., 2009).

\begin{figure}[h]
\centering
\includegraphics[height=2.5in,width=4.50in]{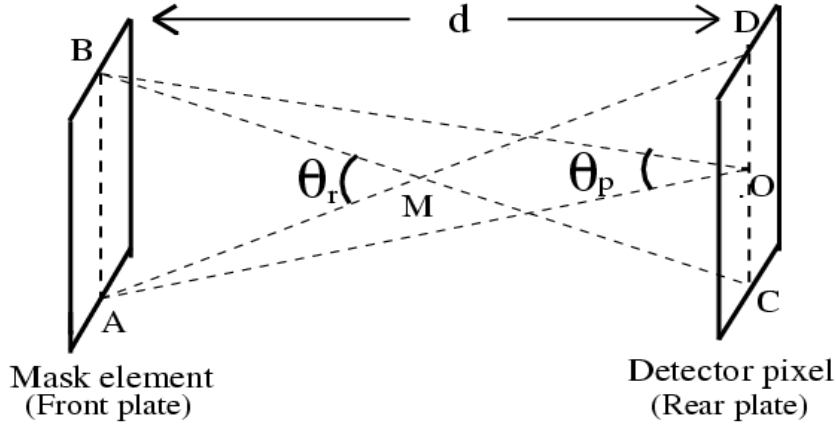}
\caption{Schematic diagram to compute angular resolution of CAM and FZP coder. In the figure, AB and 
CD are the size (width) of one single CAM element (finest zone in FZP) and detector pixel respectively.}
\end{figure}

To calculate the field of view (FOV), we use the same kind of picture as shown in Figure 3. But for
different configurations containing CAM and CZT detector, we have to replace AB and CD by an edge of 
CAM and the detector respectively and for configurations containing FZPs, we have to replace them by 
the diameters of the zone plate (FZP1 \& FZP2) far from detector and the zone plate (FZP1 \& FZP2) 
nearer to the detector respectively. 
Now, the angle subtended by AB at the point O will be the FOV of the different configurations of the 
instrument.

For such coders,  there is an inverse relationship between the angular resolution and the
FOV. Since the total height of the coding device is fixed (due to satellite
constraints), some maneuverability was available for the FZP in terms of the 
spacing between the two FZP coders. 
 If we decrease the spacing between the two zone plates, the FOV will 
increase but resolution would be poor. In our instruments, we have maintained a balance between them.
%For the FZP coders, there is a inverse relationship between the angular resolution and that of the 
%FOV. If we increase the spacing between the two zone plates, the FOV will decrease but there 
%will be better resolution. In our instruments, we have maintained a balance between them. 

In the RT-2/CZT payload, there are four configurations in which two different CAMs are used 
with two identical CZT detector modules, whereas for other two configurations dual FZPs are used 
with CZT and CMOS detectors respectively. The configuration details are given in Table 3.

CONFIG-1 contains CAM1 as the coder and CZT1 as detector. Smallest coder element size (AB) is 
of $0.25~cm$ and spacing between CAM1 and CZT1 is $40~cm$. So the angular resolution comes out to be 
$21.5~arcmin$ and FOV is 5.72$^\circ$. 

CONFIG-2 also contain a CAM (CAM2) and a CZT detector (CZT2). In this case, as the parameters are 
the same as that of previous one the angular resolution and FOV have the same value as the previous 
configuration (CONFIG-1).

In CONFIG-3, the coder is dual FZP (FZP1) with diameter $3.0~cm$ and the
finest zone width is $0.0050~cm$. The spacing between the two zone plates is $32~cm$. The 
angular resolution comes out to be $64~arcsec$.
Unlike the other cases, in FZP configurations, the FOV is not determined by the diameter and 
spacing only. Finite size of the detector pixel actually put limitation on the FOV (Chakrabarti et 
al., 2009 \& Palit et al., 2009). As fringe separation decreases with an increase in 
the  off-axis angle of 
the source, the detector pixel can not differentiate two adjacent fringes if the separation between 
them goes below the detector pixel size for any large off-axis source. Hence, reconstruction of 
those sources  is not possible. Detector pixel limited FOV for this configuration (CONFIG-3) is 
found to be $409~arcsec$ wide.   

CONFIG-4 is designed with dual FZP (FZP2) as shadow-caster of diameter $2.4~cm$ and a high spatial 
resolution CMOS as a detector. The width of outermost zone is $0.0041~cm$ and the spacing 
between two FZP is $32~cm$. The maximum achievable angular resolution is around $54~arcsec$ and FOV is 
4.29$^\circ$.

In the present configuration, the FZP coders have superior angular 
resolution compared to CAM.
%are far superior to the CAMs in terms of angular resolution.
In the case of FZP with CZT detector (CONFIG-3) configuration, the FOV is limited by large pixel 
size of detector to a very small value. 
%So, it is not possible to do an extensive observation with 
%such combination. 
The configuration of CMOS detector with FZP is the  most viable option to image the hard 
X-ray solar flares. 
 
%In the next section, we have carried out extensive simulations for all the four configurations to see 
%the expected feasible results, which could be achievable from  the RT-2/CZT payload.

\section{Simulations}

All the shadow casters along with the detectors, when placed in orbit to grab images of the source,  
receive parallel rays of radiation as the sources are effectively at infinite distance. 
In general,
it is difficult during laboratory experiments to have a source which is at an infinite distance so that the 
shadow caster can receive parallel beam of X-rays. 
%The use of collimator for generating parallel sources is not basically possible for high energy photon 
%beam such as hard X-ray. 
For experimental arrangements however large the X-ray source distance is made, the 
effect of divergence of the beam appears in the reconstructed source figures as can be seen in the 
laboratory results corresponding to the FZP coders (see  $\S$6). 
Though with some modifications in one of the coder plates such effects
can be rectified (Palit et al., 2009), 
with the present configuration (employed in RT-2/CZT payload) of the zone plates
effect of the divergent beams cannot be eliminated.
Further, it is also difficult to have more than
one hard X-ray source to study the properties of 
angular resolution and FOV. One of the ways to understand the characteristics 
of the coders and to examine their efficiency in getting images 
is to simulate
extensively all the required cases for all four configurations keeping all the parameters and 
environment intact. 
 
Simulations are done for every configuration and for all cases involving varying number of 
sources, with large off-axis angle of source for the  verification of FOVs and for closely 
placed sources to verify the mathematically obtained angular resolutions. 
The infalling photon number on the front coders (CAM1, CAM2, FZP1 and FZP2) of all four configurations
is chosen to be 5$\times$$10^{5}$, while considering a single source. 
For double or multiple sources with varying intensity, photon numbers for the brightest source remains 
the same and photon numbers for relatively lower intensity sources are mentioned in respective sections.
This number is sufficient for hundred second (onboard accumulation time for each frame) data 
accumulation by the imager (4 different configuration for imaging in RT-2/CZT payload) from a C class 
flare (and above) that occur on the surface of the Sun.
 
%In all the following 3D plotting of the reconstructed source planes, the scalings for the 
%coordinate axis corresponding to the strength of reconstructed sources are arbitrary and program 
%generated. But with the reference from those figures we get estimations of the relative strengths of 
%sources during real time data analysis. All simulation results are given below.

Simulation results are interpreted based on the 2D and 3D representations of the reconstructed source
position and relative strength of the peaks. Source intensity variation in reconstructed image plane is 
plotted in arbitrary units. Detailed simulation results for all four configurations are presented in the
following sections.
 
\subsection{CONFIG-1: CAM1 + CZT1} 

CONFIG-1 consists of a single coder CAM (CAM1) and a CZT (CZT1) module and both are placed 40 cm apart 
in the first quadrant of the collimator. CAM pattern for this configuration is shown in Figure 4(a) 
(top left). The CAM pattern 
is generated from the first polynomial given in section 2.1.1. A source position is generated at
a position of $\theta$ = 42$^\circ$ and $\phi$=$1^\circ 47'$. 
The angle $\theta$ is  measured taking positive horizontal
axis as polar $x$ coordinate ($\theta$=$0^\circ$) and 
$\phi$ is the angle from the vertical axis.
The shadow of CAM due to this
source on detector plane is shown in Figure 4(b) (top right).
The shift of shadow pattern of the CAM from central position has coded the information on 
the position of source with respect to the central point in FOV. Reconstruction of the image 
(source position) from the CAM pattern is done according to the method discussed in section 2.1.2.
In Figure 4(c)(bottom left) and Figure 4(d) (bottom right), we have shown the 2D and 3D view of 
the reconstructed source (image). In both the figures of the reconstructed sky plane, the FOV is 
5.72$^\circ$ wide along each sides. The reconstructed source position as we evaluated from 
Figure 4(c), matches with the actual source position assigned during simulation. 
From Figure 4(d), we also get a measure of the actual intensity of the source.
 
\begin{figure}[h]
\centering
\includegraphics[height=1.8in,width=1.8in]{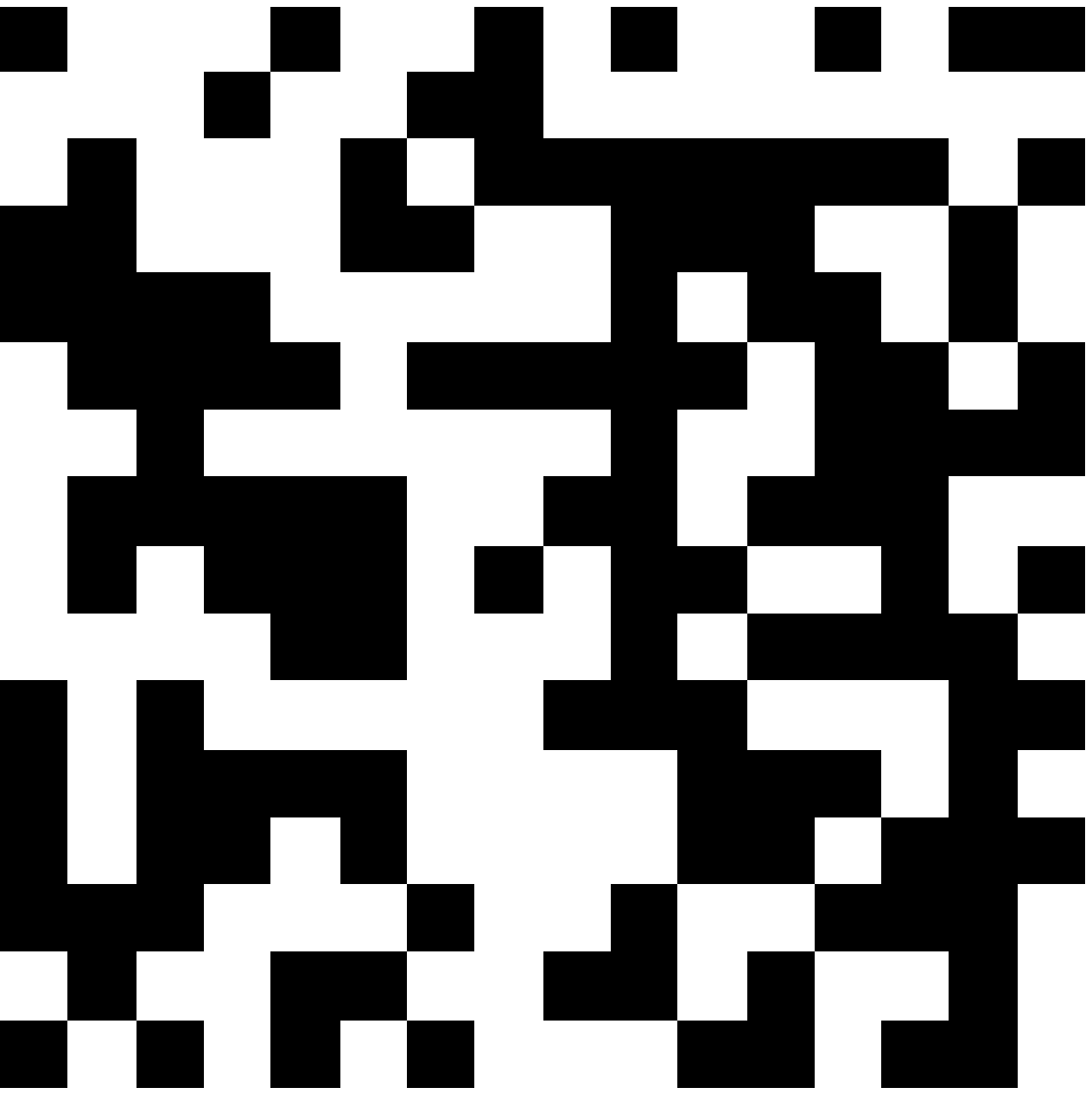}\hspace{1.0 cm} %1
\includegraphics[height=1.8in,width=1.8in]{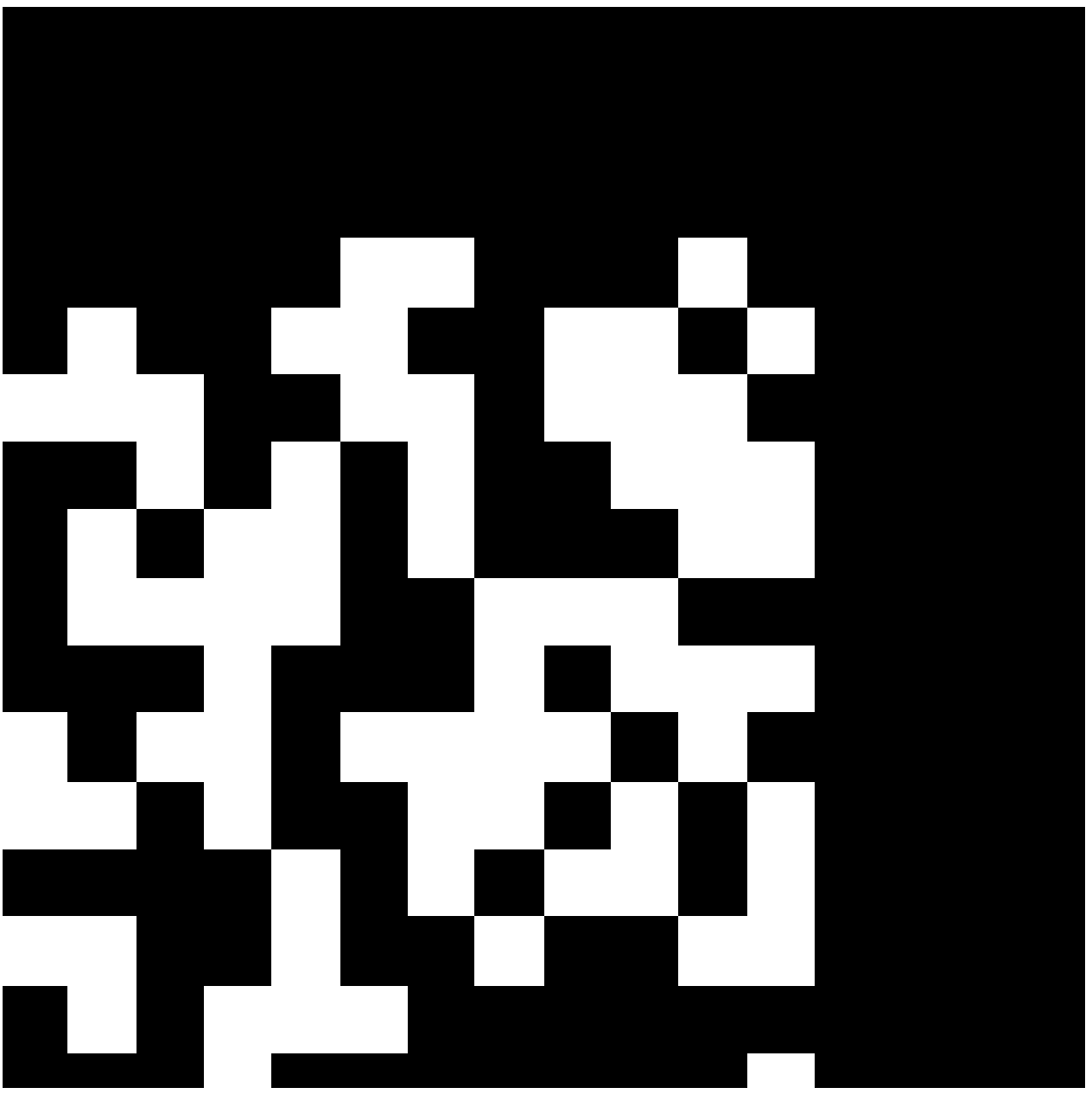}
\includegraphics[height=1.8in,width=1.8in]{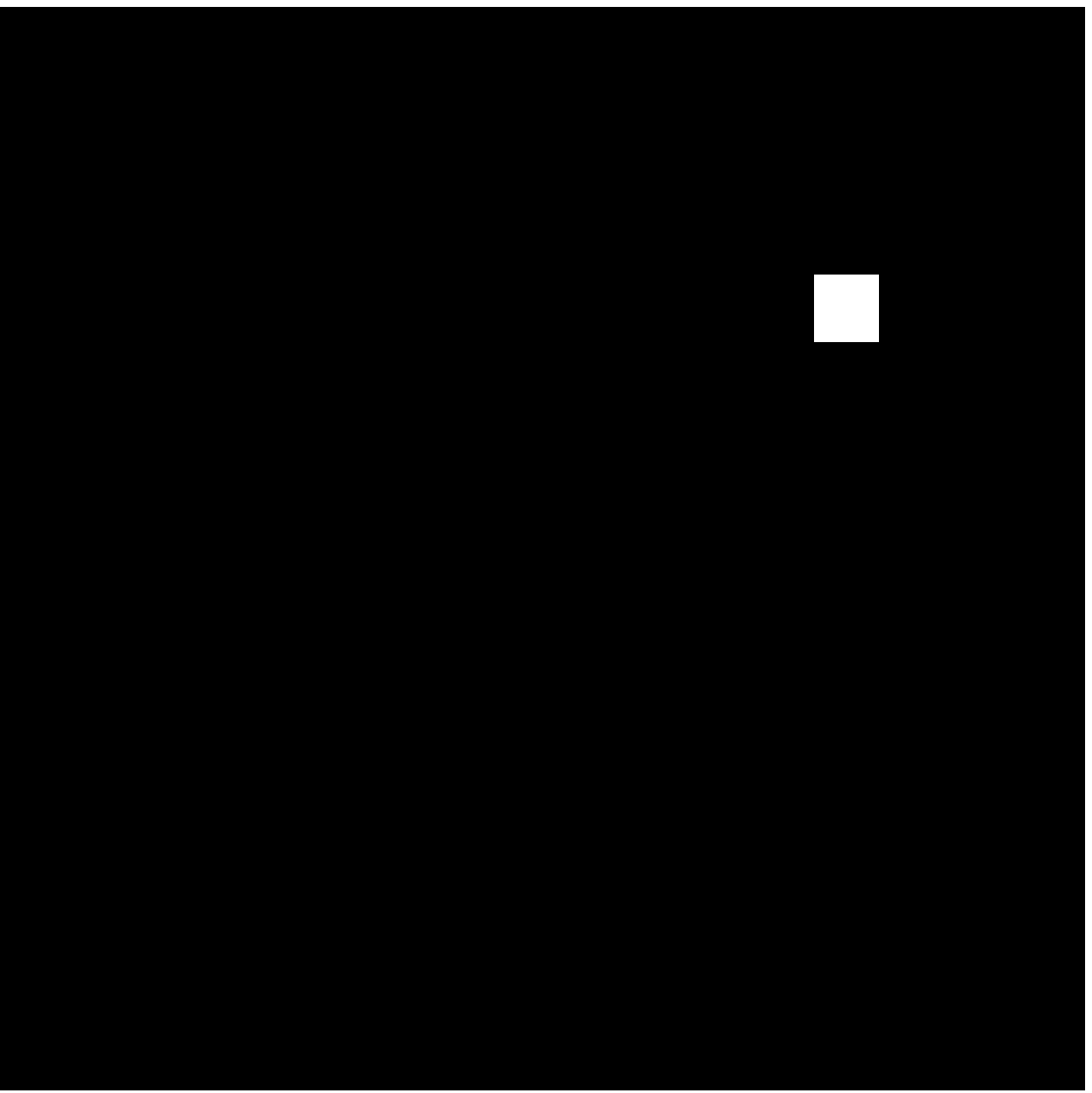}
\includegraphics[height=2.0in,width=2.4in]{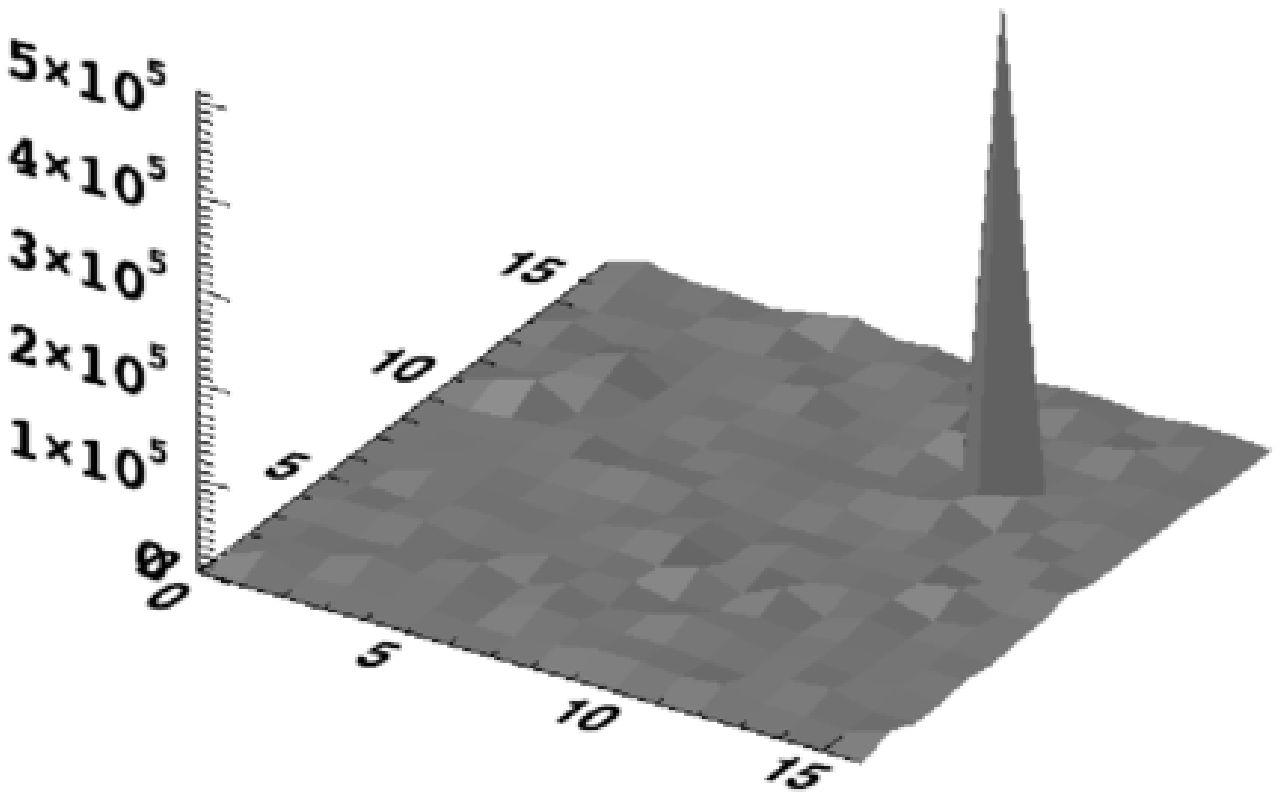}
\caption{(a) Simulated picture of CAM1 pattern, where opaque elements are coded by black color. 
(b) Shadow pattern obtained in the CZT 
detector for CONFIG-1 for a single source. (c) Two dimensional view of reconstructed sky plane. 
(d) Three dimensional picture of the source plane obtained by reconstruction.}
\end{figure}

In Figure 5(a-c), we present the simulation results to verify the accuracy of 
the mathematically obtained angular resolution of the configuration containing CAM1 and CZT1. For 
this, we have placed two sources at an angular separation of 21.5 arc-minute from each other. The 
number of photons falling on the CAM from the brighter source is 5$\times$$10^{5}$ and that from the 
fainter one is 2.5$\times$$10^{5}$. In Figure 5(a), we have shown the shadow pattern obtained for two
sources which are placed very close to each other. The reconstructed source (image) in 2D and 3D 
view of both the sources are shown in Figure 5(b)(top right) and 5(c)(bottom). It can be seen from the 
figures that the two sources are separated by one pixel between them. So the sources can be said to be 
just resolved. 
  
\begin{figure}[h]
\centering
\includegraphics[height=1.8in,width=1.8in]{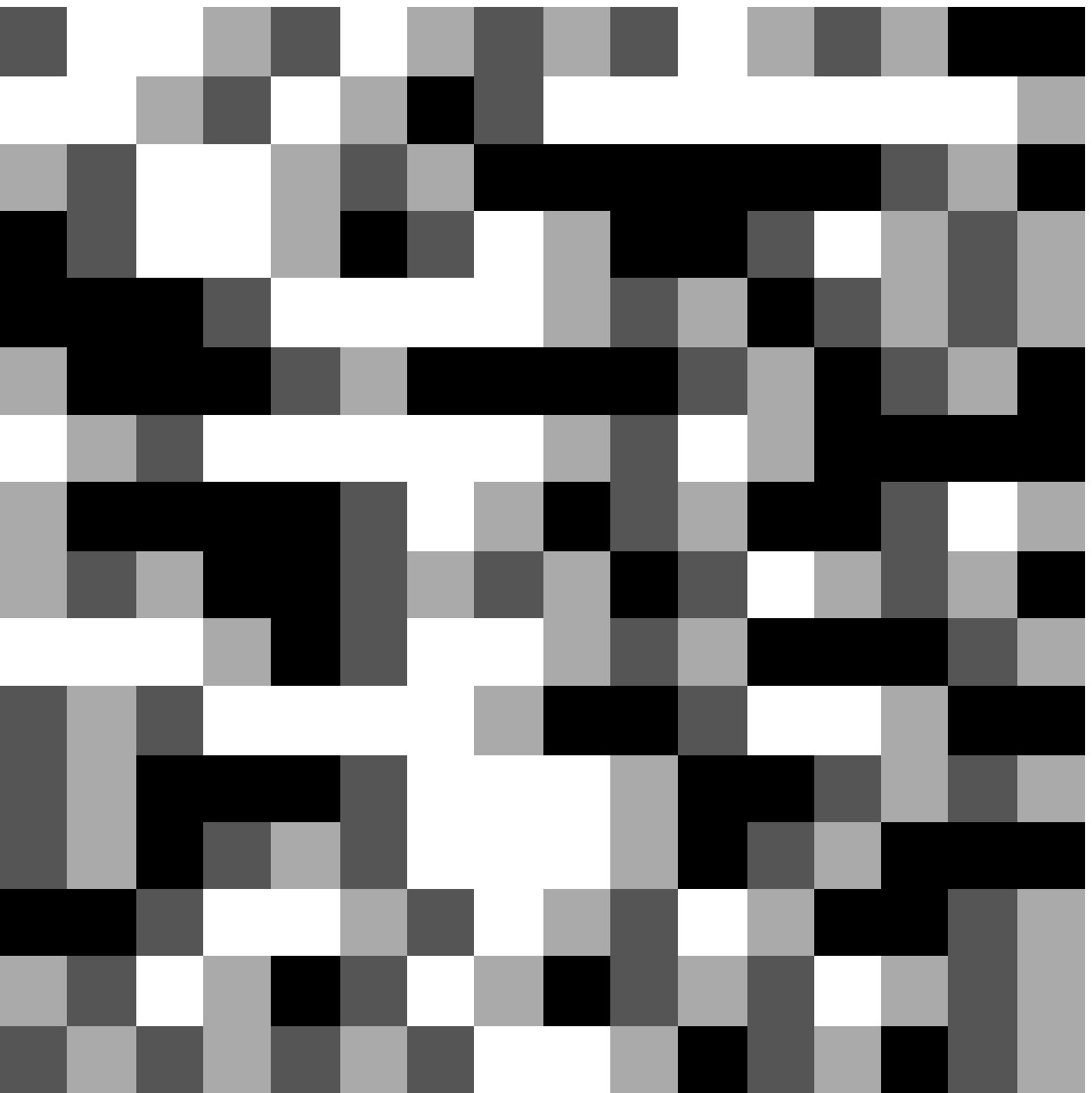}\hspace{1.0 cm} %2
\includegraphics[height=1.8in,width=1.8in]{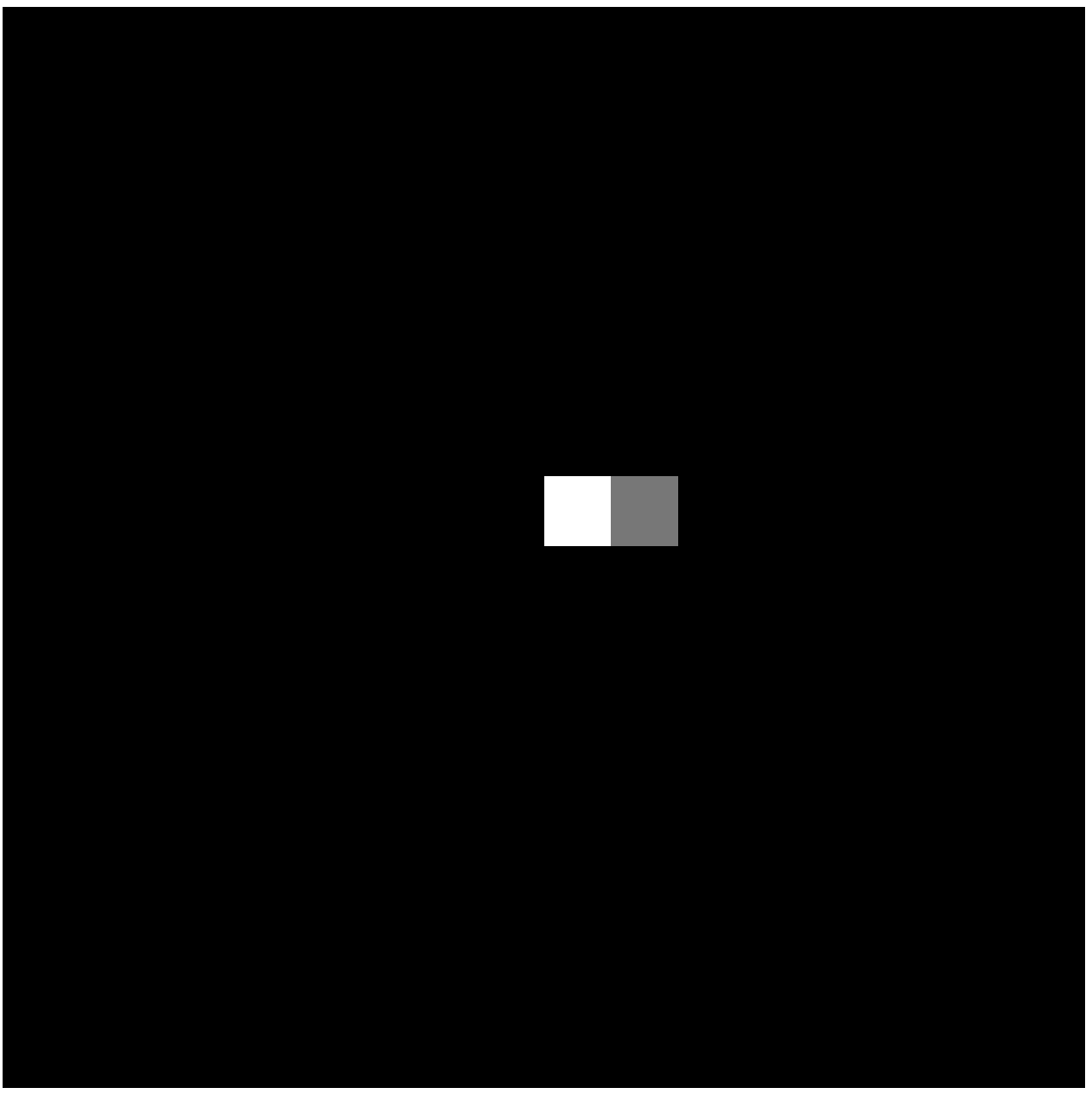}
\includegraphics[height=2.8in,width=3.4in]{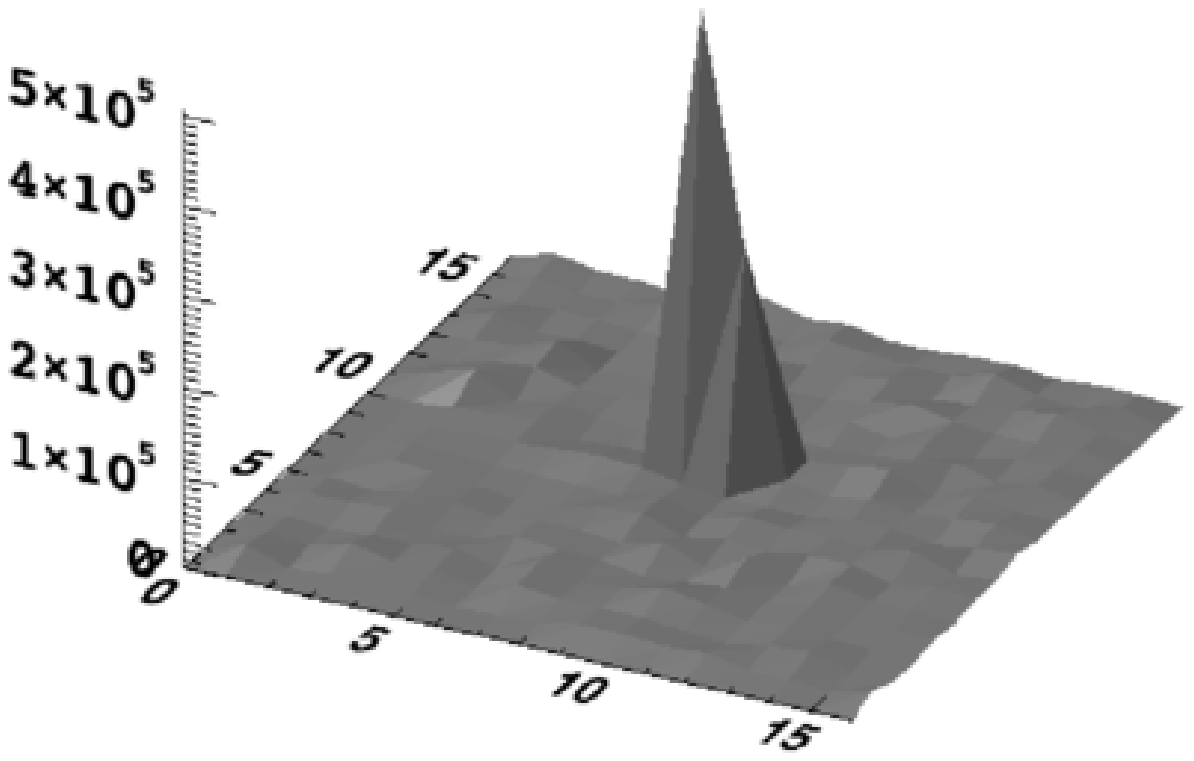}
\caption{(a) Shadow pattern obtained for two sources placed very close to each other 
(21.5 arc-min) with CAM1 
and CZT1 (top left) configuration. (b) 2D view of reconstructed sky plane (top right). (c) 3D
view of the source intensities obtained by reconstruction (bottom).}
\end{figure}

The CONFIG-1 and CONFIG-2 are identical as both configurations use CAM and CZT except the CAM 
patterns are different in CAM1 and CAM2. Therefore, the FOV and angular resolution in both 
configuration are mathematically the same. 

\subsection{CONFIG-2: CAM2 + CZT2}

The CAM used in this configuration is different from the CAM pattern of CONFIG-1. The CAM2 pattern is 
generated from the second polynomial given in section 2.1.1 and is shown in Figure 6a
(top left). Simulation results for CONFIG-2 (CAM2 and CZT2) are shown in Figure 
6(b,c,d). In this case, we have considered two sources in the field of view of the collimator with 
position $\theta$ = 227$^\circ$ $30'$, $\phi$=$3^\circ$ $00'$ and $\theta$ = 90$^\circ$ $00'$, 
$\phi$=$00^\circ$ $32'$ respectively. 
%The angle $\theta$'s are measured taking positive horizontal
%axis as polar $x$ coordinate ($\theta$=$0^\circ$). 
Number of X-ray photons falling on CAM2 
from the two sources are the same (5$\times$$10^{5}$ counts).
In Figure 6b (top right), we have presented the combined shadow pattern of the CAM (CAM2) generated by 
the two sources. 

\begin{figure}[h]
\centering
\includegraphics[height=1.8in,width=1.8in]{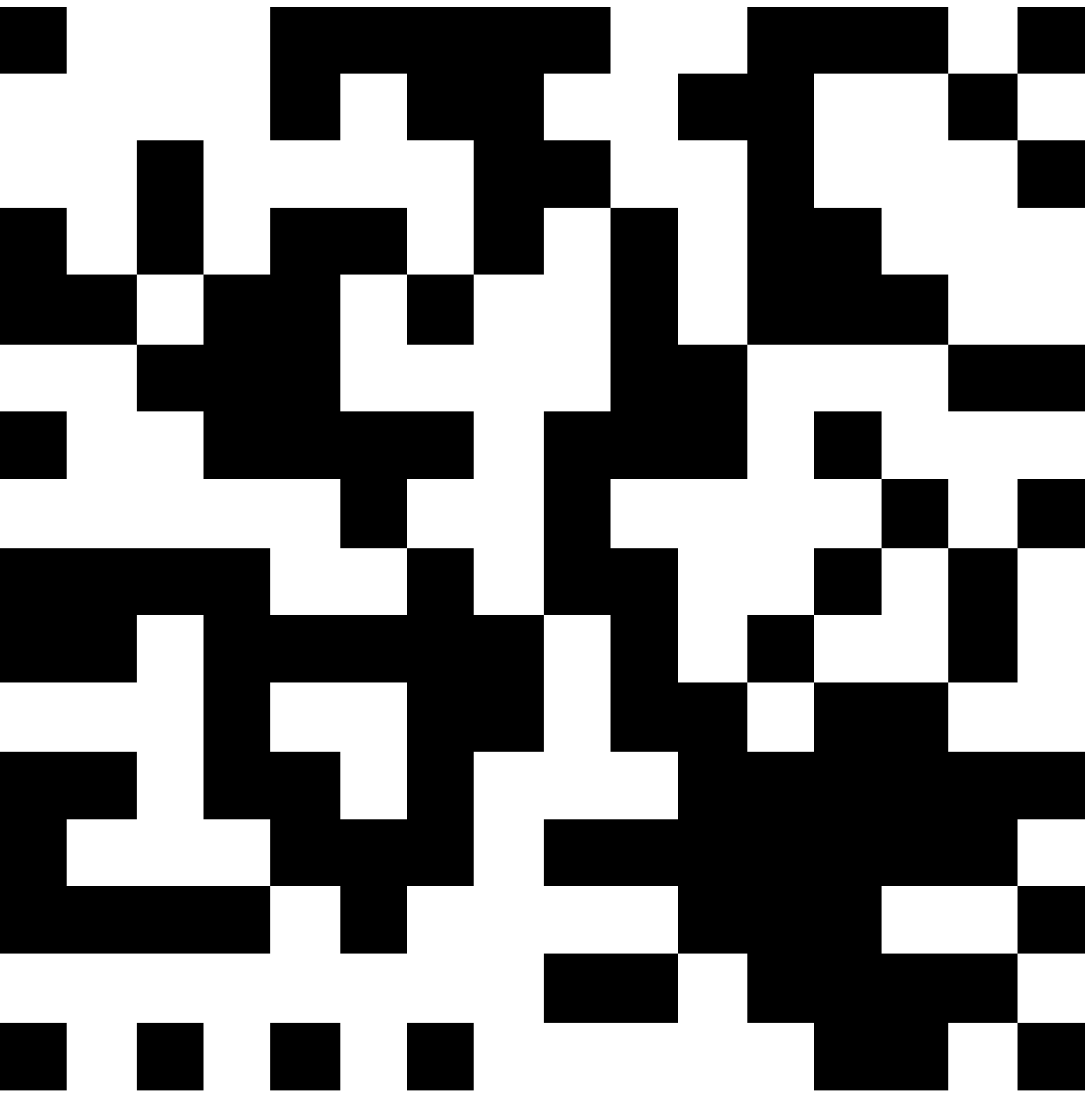}\hspace{1.0 cm} %3
\includegraphics[height=1.8in,width=1.8in]{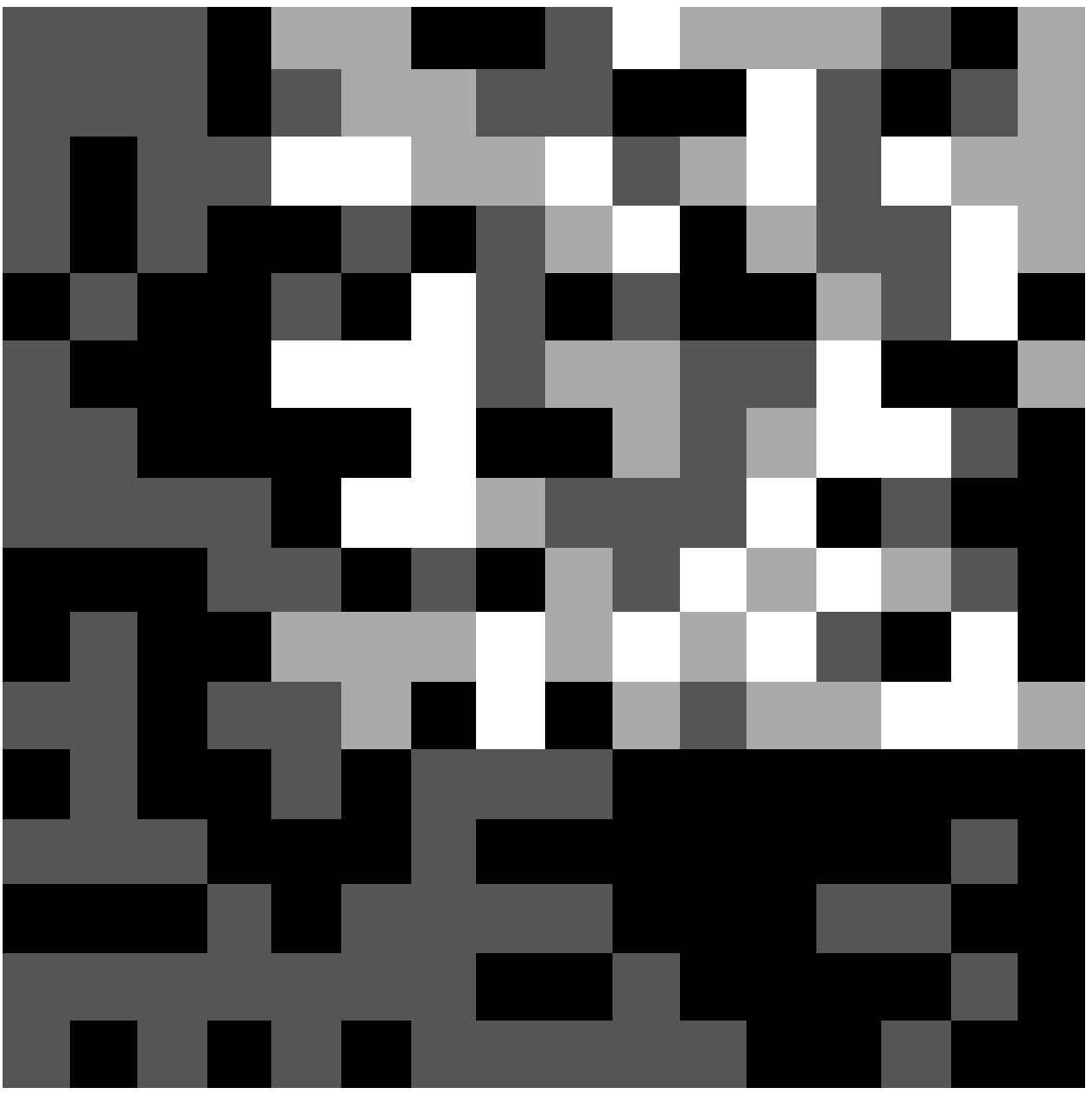}
\includegraphics[height=1.8in,width=1.8in]{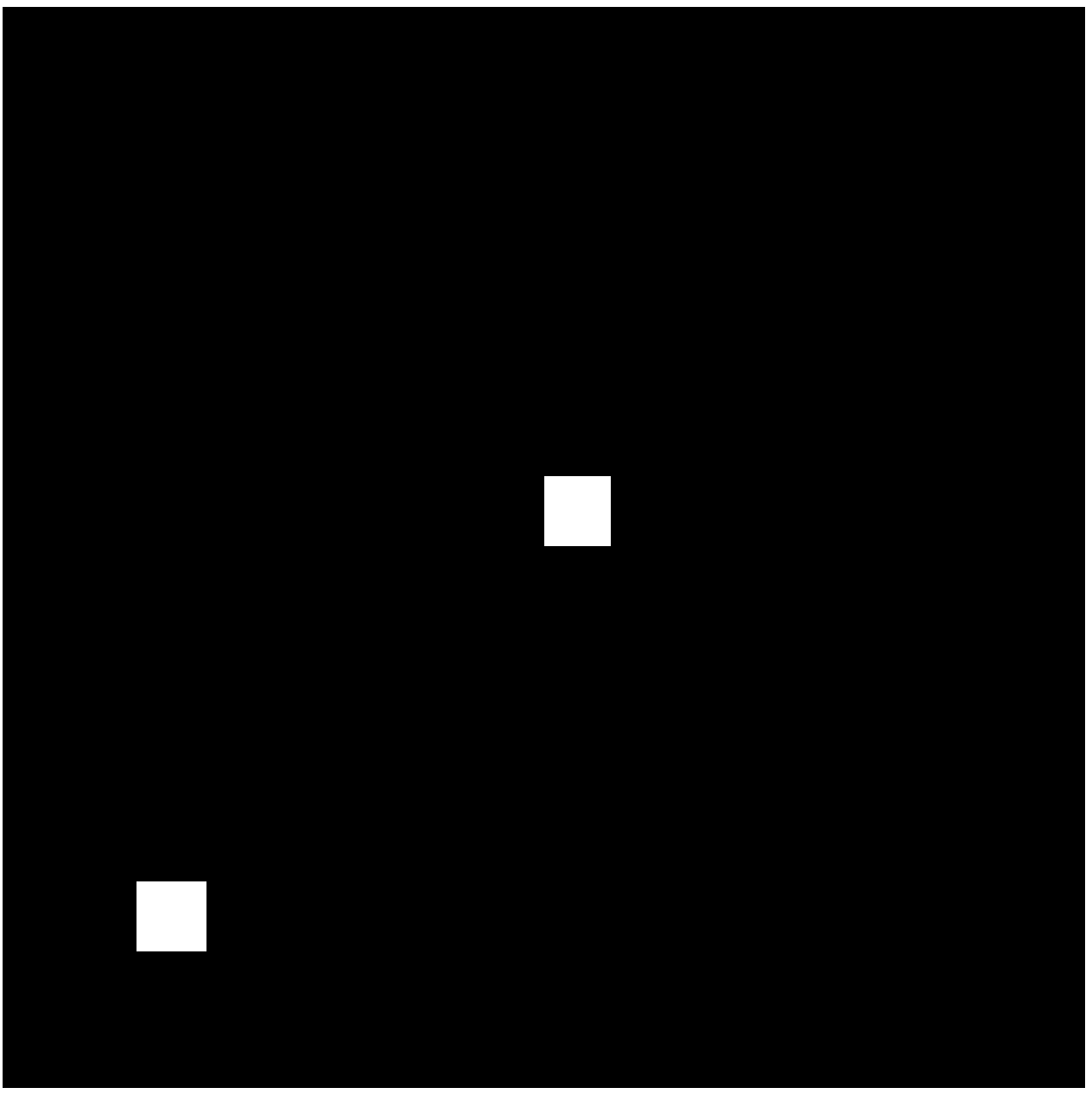}
\includegraphics[height=2.2in,width=2.4in]{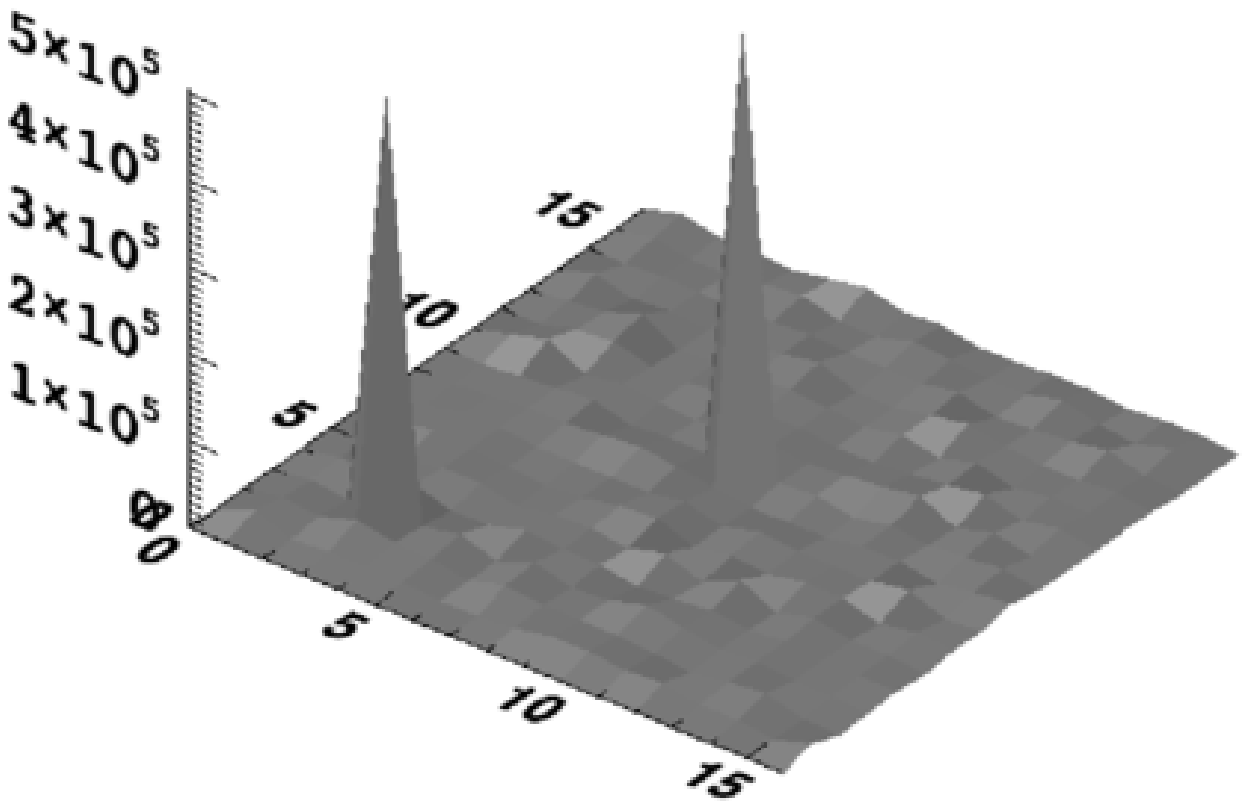}
\caption{Simulated picture (opaque elements of the pattern are coded by black color) of CAM used in 
CONFIG-2 (top left). (b) Shadow pattern obtained 
for two sources (top right). (c) 2D view of reconstructed sky plane (bottom left). (d) 3D picture 
of the sources obtained by reconstruction (bottom right). See text for details.}
\end{figure}
 
The shifts of individual shadows produced by each source is compatible with their positions as 
can be seen from the Figure 6. In Figure 6c (bottom left), we have presented the 2D picture of the 
reconstructed object plane. From this figure, we get the exact information about the positions of 
the two sources. These positions agree with the assigned positions of the sources during simulation. 
From Figure 6d (bottom right), which is the 3D view of the reconstructed source plane,  
we get the information on the relative brightnesses of the sources. The two 
peaks correspond to the two reconstructed sources and they are found to be of the same height (intensity). It is 
clearly seen from this simulation that the relative brightnesses of the sources in reconstructed plane 
are exactly replicated irrespective of their relative positions whereas in case of many other imaging 
devices there may be position dependencies of relative intensities of the reconstructed sources.   

\begin{figure}[h]
\centering
\includegraphics[height=1.8in,width=1.8in]{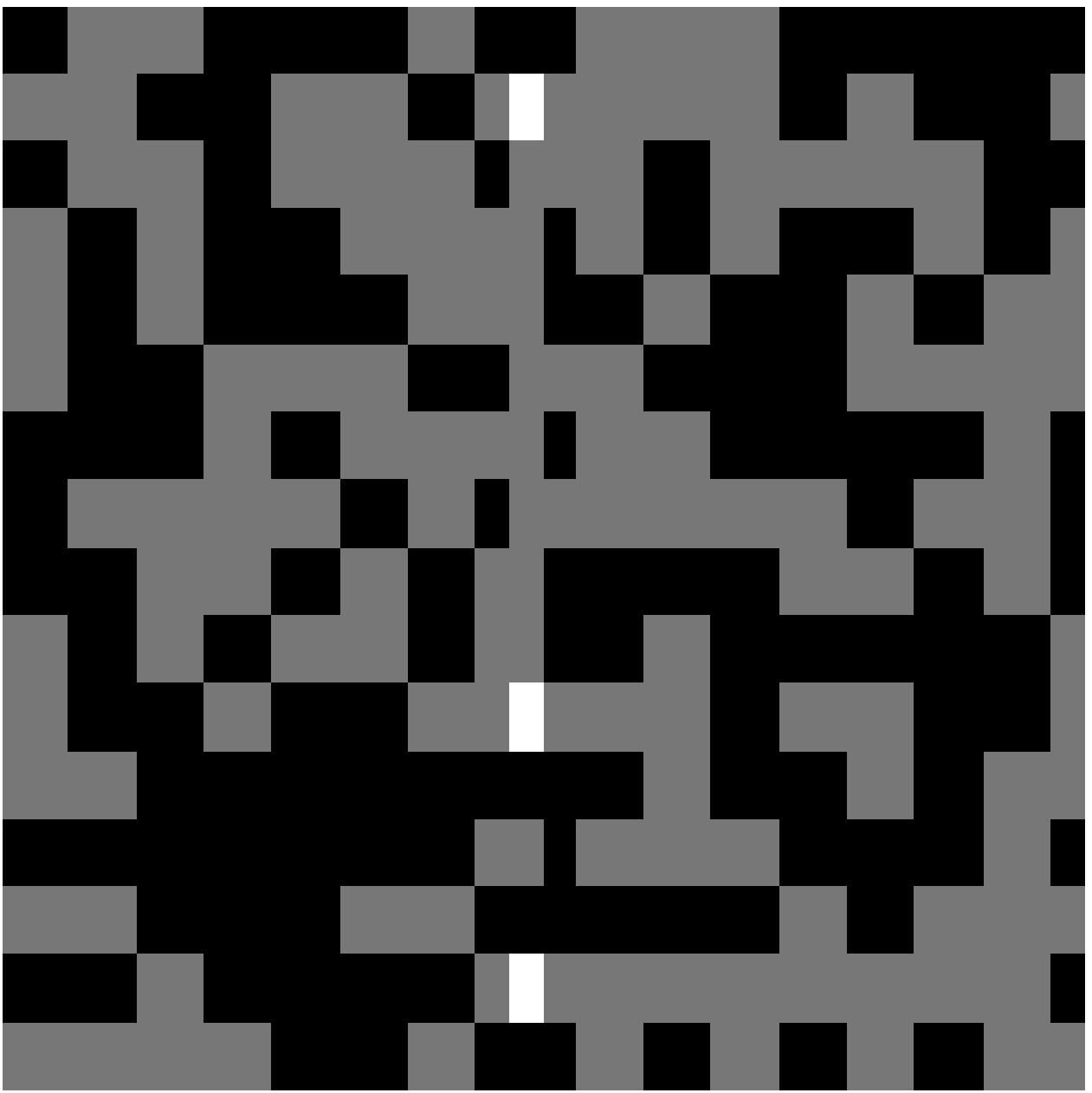}\hspace{1.0 cm} %4
\includegraphics[height=1.8in,width=1.8in]{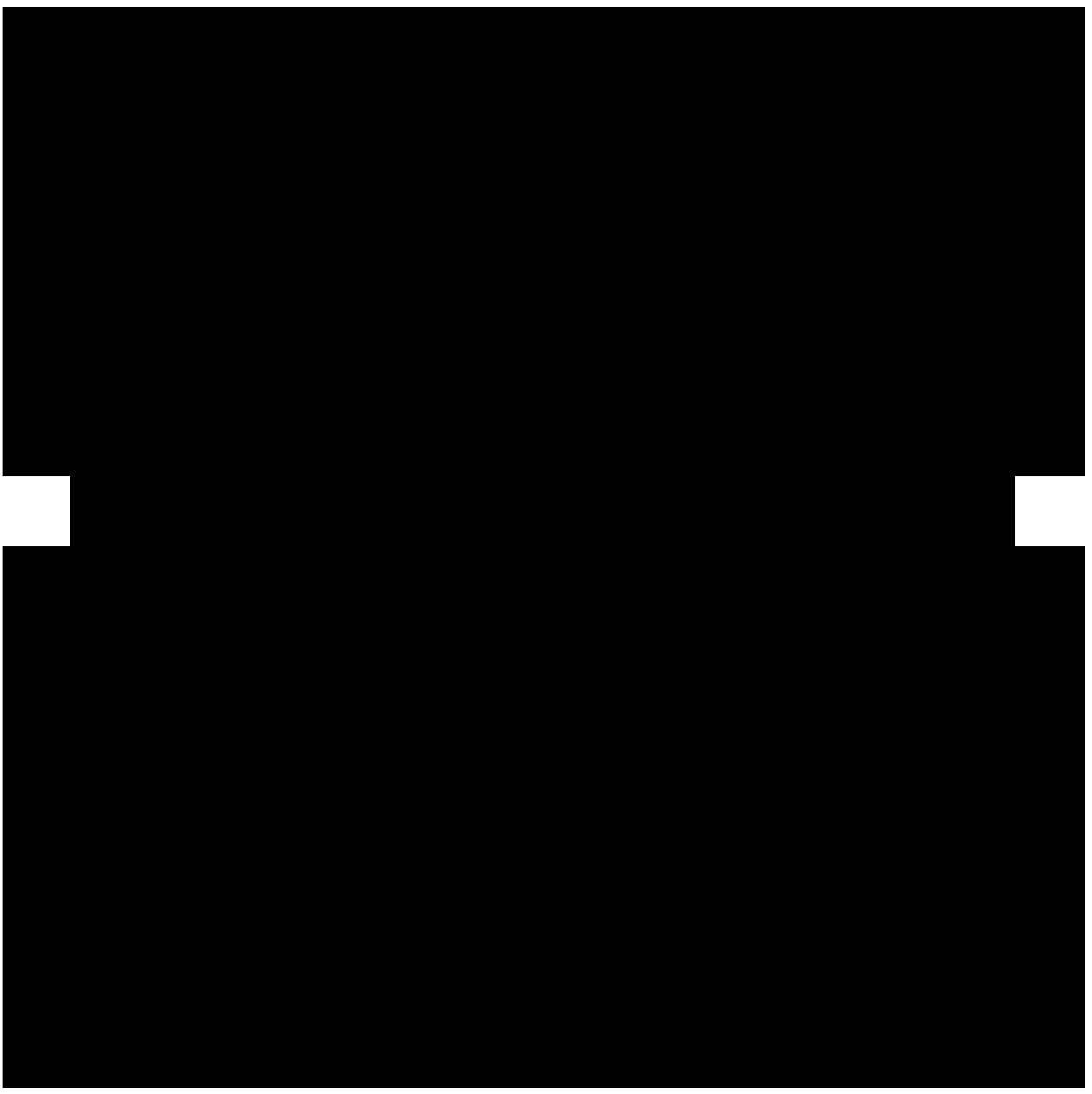}
\includegraphics[height=2.8in,width=3.3in]{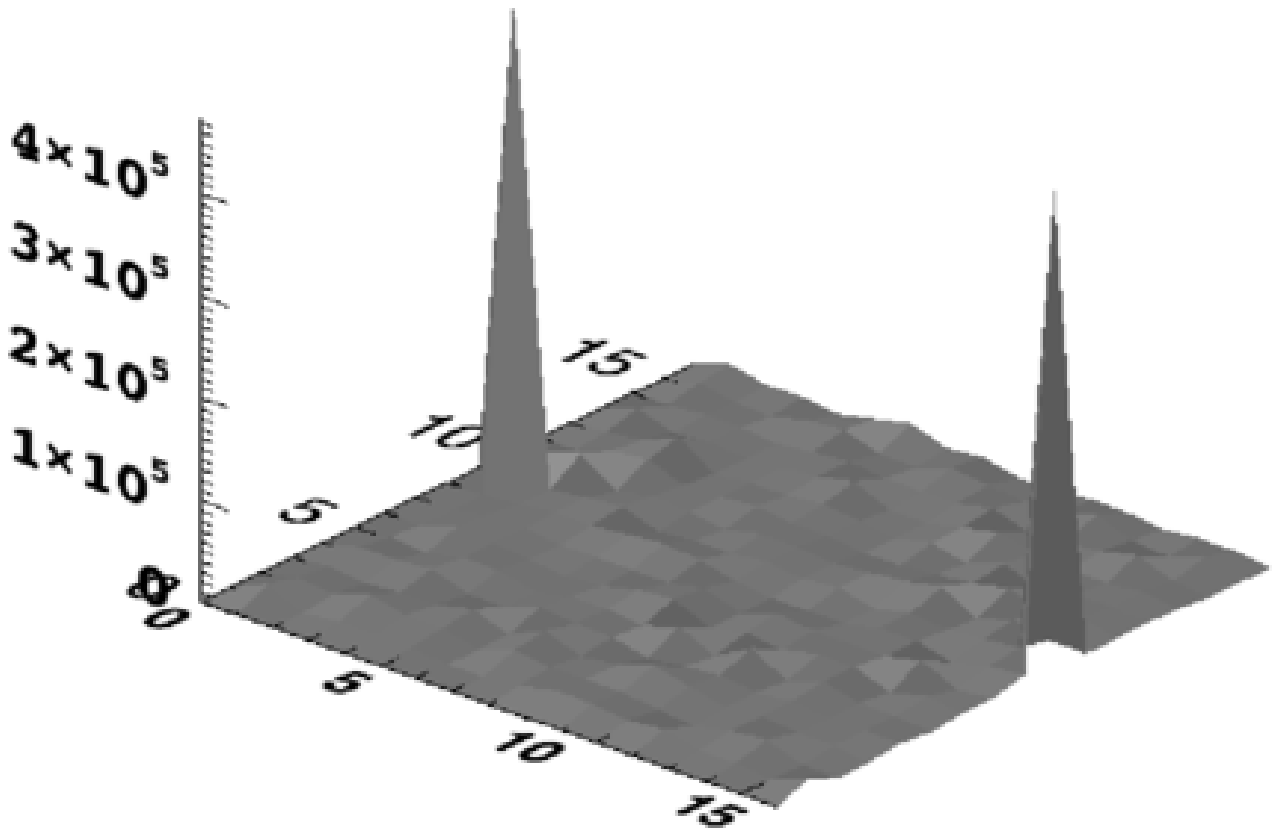}
\caption{(a) Shadow pattern of CAM2 for two sources placed at two extreme ends of FOV (top left),
(b) 2D view of reconstructed sky plane (top right).(d) 3D view of the sources obtained by 
reconstruction.}
\end{figure}

In the next simulation step, we place two sources at extreme right and left ends of the 
collimator wall to verify mathematically calculated FOV. 
Sources are placed at $\theta$ = 0$^\circ$ and  $\theta$ = 180$^\circ$ with common $\phi$ angle 
of value 2.86$^\circ$. It is clearly seen from the reconstructed images that the sources are really 
placed at the two extreme edges of the collimator and their separation is the measured
value of FOV. In Figure 7(b, c), 2D and 3D view of the reconstructed images of the sky plane with 
two sources are shown.

\subsection{CONFIG-3: FZP1 + CZT3}

The CONFIG-3 consists of a pair of Fresnel zone plates (FZP1) as a 
coder and a CZT module (CZT3) as detector. Both the FZPs have inner radii of 0.122 cm and 
the number of zones in each is 151. The pair is a positive cosine one, i.e, the $n^{th}$ zone radius
for each of the zone plates is equal to $\sqrt n$ times the inner radius and the central zones are
transparent to X-rays. CZT detector consists of 256 pixels having dimension of 0.25 cm $\times$ 
0.25 cm.

\begin{figure}[h]
\centering
\includegraphics[height=1.8in,width=1.8in]{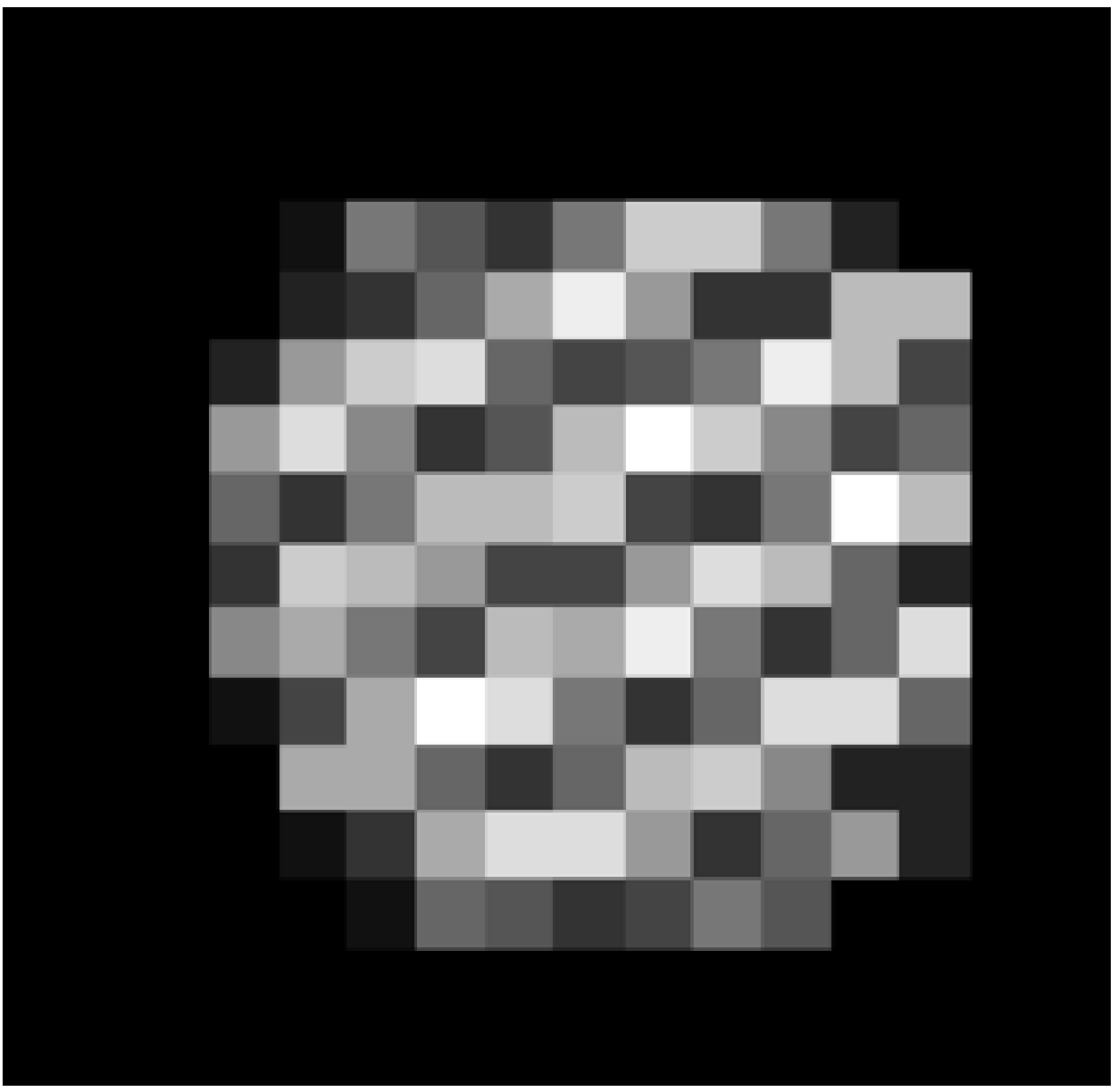}\hspace{1.0 cm} %5
\includegraphics[height=1.8in,width=1.8in]{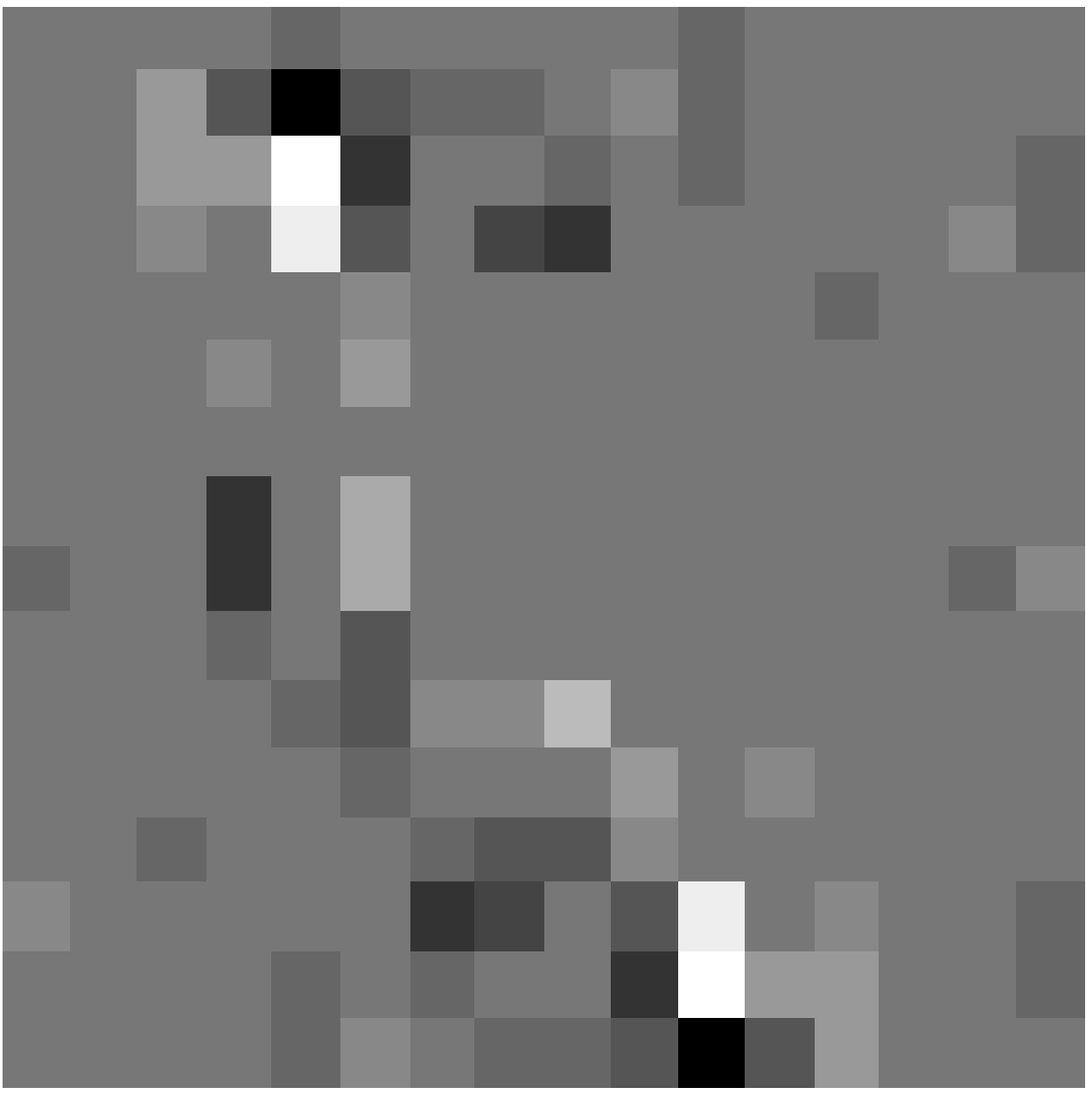}
\includegraphics[height=3.0in,width=3.0in]{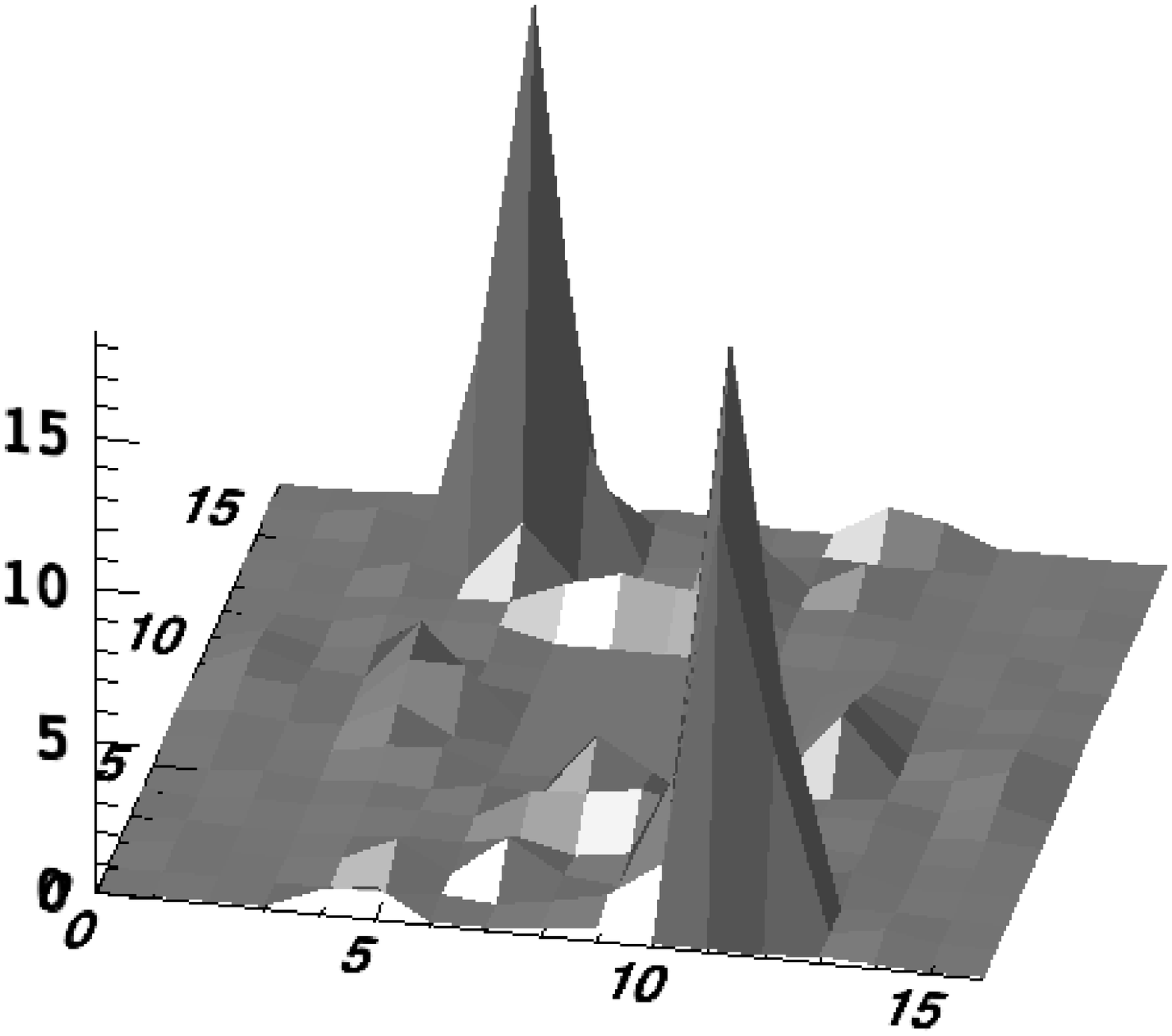}
\caption{(a) Fringes obtained on CZT3 detector with FZP1 as coder
(top left). Two sources appear in the reconstructed 2D (b) and 3D (c) image plane, one of
which is the pseudo source of the actual source. The central DC-offset is chopped out.}
\end{figure}

In Figures 8(a-c), we have plotted the fringe and reconstructed sources for a single source placed 
at $\theta$ =116$^\circ$ and $\phi$ = $170''$. We have chopped out the central DC-offset 
(Chakrabarti et al. 2009) to get the prominent source picture, while plotting the reconstructed images. 
A pseudo source apart from the original source
position is seen in the reconstructed image. Combination of cosine and sine FZPs can remove the
effect of ghost image (pseudo source). The finite size of the detector pixel restricts the field of 
view to $409''$.

FOV calculated for this configuration can be verified from the simulation too. For the 
simulation,
we have considered a single source placed at $\theta$ = 0$^\circ$ and $\phi$ = $204''$.

\begin{figure}[h]
\centering
\includegraphics [height=1.8in,width=1.8in]{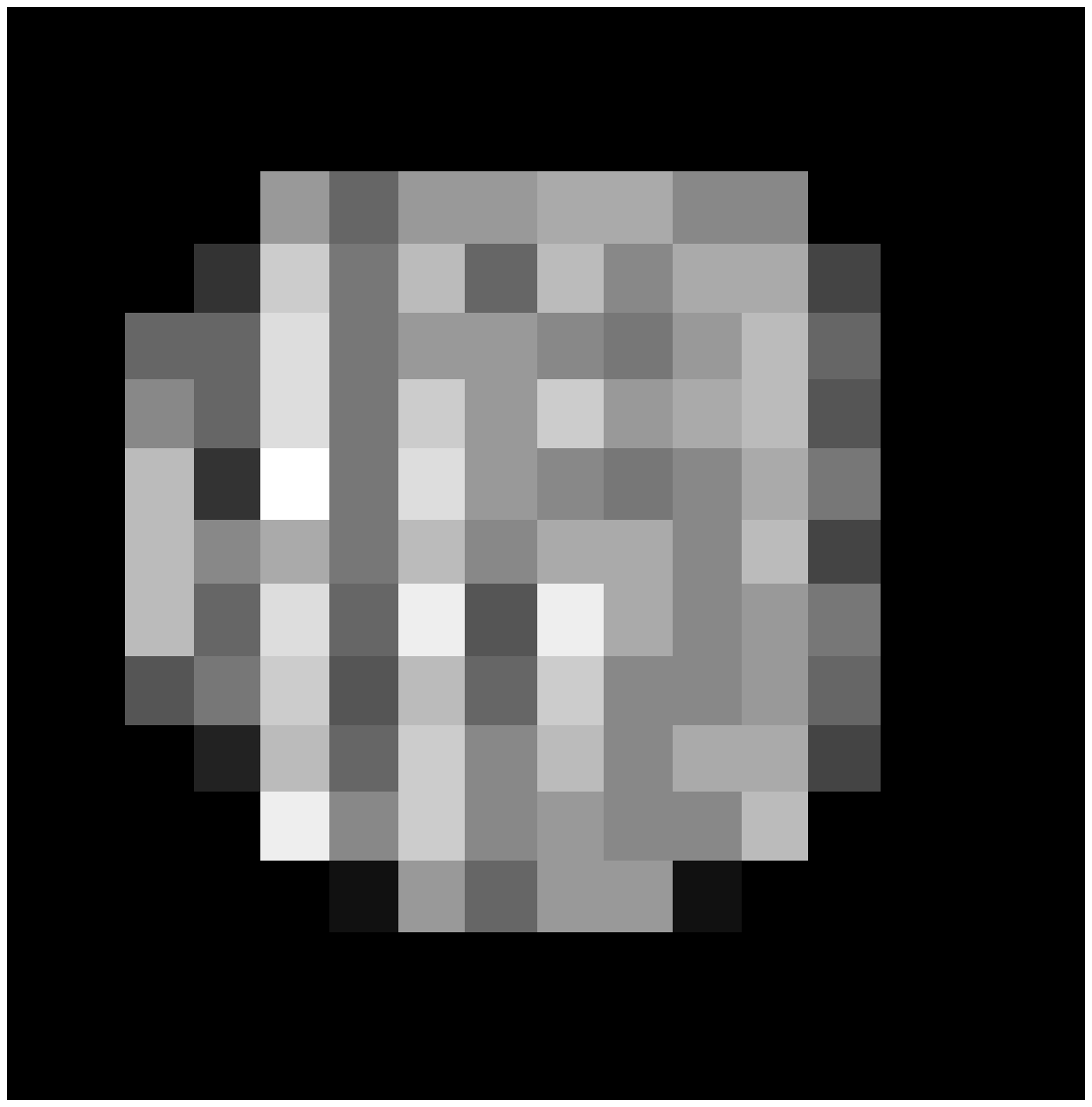}\hspace{1.0 cm} %6
\includegraphics[height=1.8in,width=1.8in]{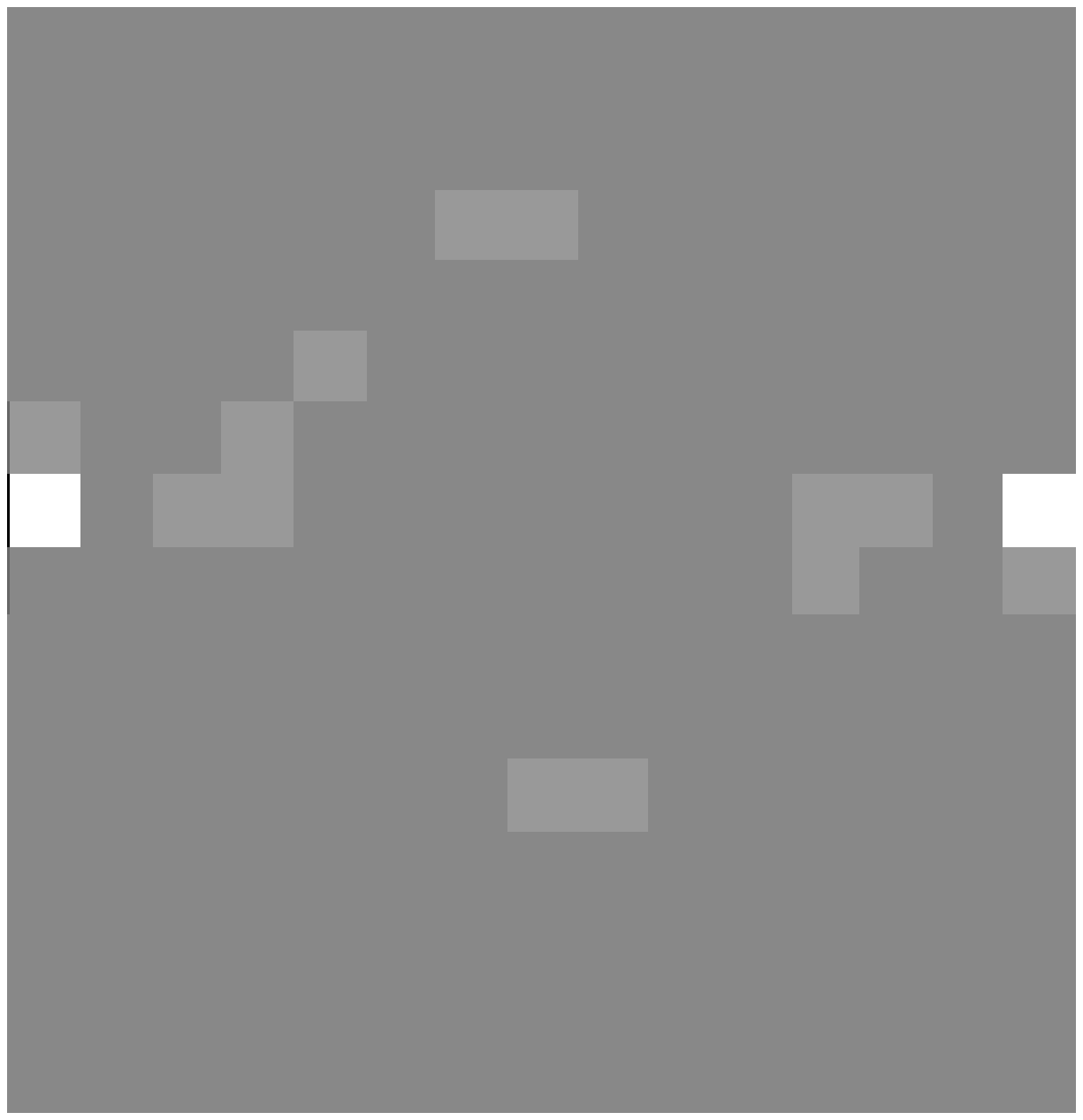}
\includegraphics[height=3.2in,width=3.2in]{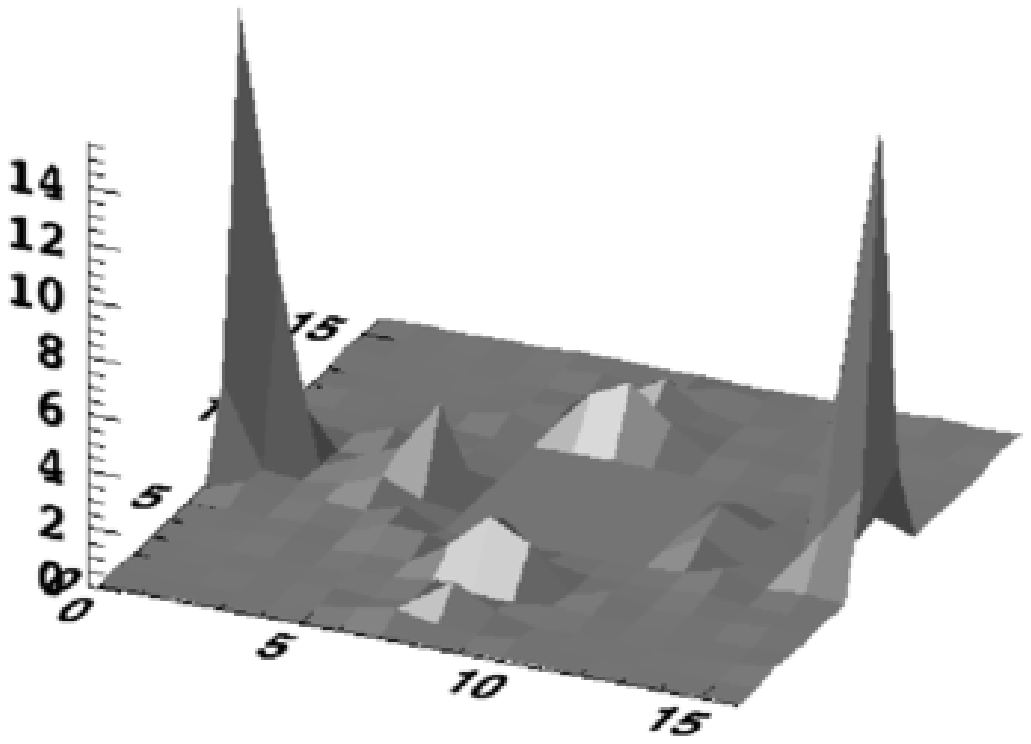}
\caption{(a) Fringes obtained on a CZT detector combined with a FZP coder (top left).
The source is at extreme right edge of the reconstructed sky array shown in the reconstructed 2D 
(top right) and 3D (bottom) view.}
\end{figure}

In Figure 9(a-c), we have plotted the fringe pattern along with 2D and 3D reconstructed source in the
detector plane. It is seen that the reconstructed source (also the pseudo source) is at 
extreme edge of the reconstructed array. This confirms that the FOV of the instrument is actually 
twice the angle of $\phi$, i.e, $409''$. 

We have also carried out simulation to verify the angular resolution achievable by this configuration.
According to the design specifications, the calculated angular resolution is around $64''$ (see
section 3). In Figures 10(a-c), we have presented the combined fringes and reconstructed 2D and 
3D views of reconstructed sky plane of two closely placed sources. The sources are placed at roughly 
64 arc-sec apart.
From the 2D (top right) and 3D (bottom) 
images of the reconstructed sources, we find that two sources are just resolved. 
Hence we can conclude 
that the mathematically found angular resolution is well supported by simulations.

\begin{figure}[h]
\centering
\includegraphics[height=2.0in,width=2.0in]{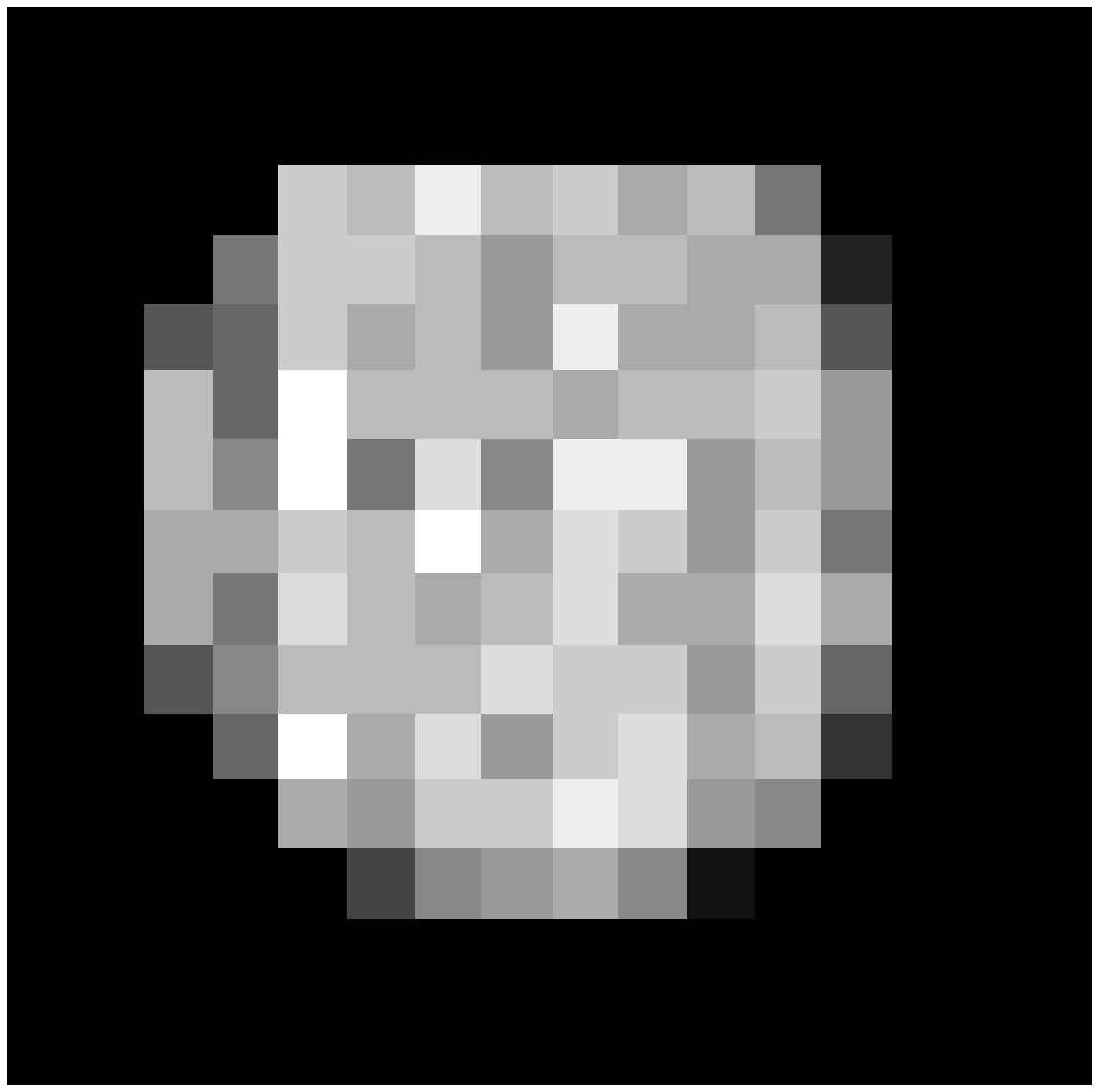}\hspace{1.0 cm} %7
\includegraphics[height=2.0in,width=2.0in]{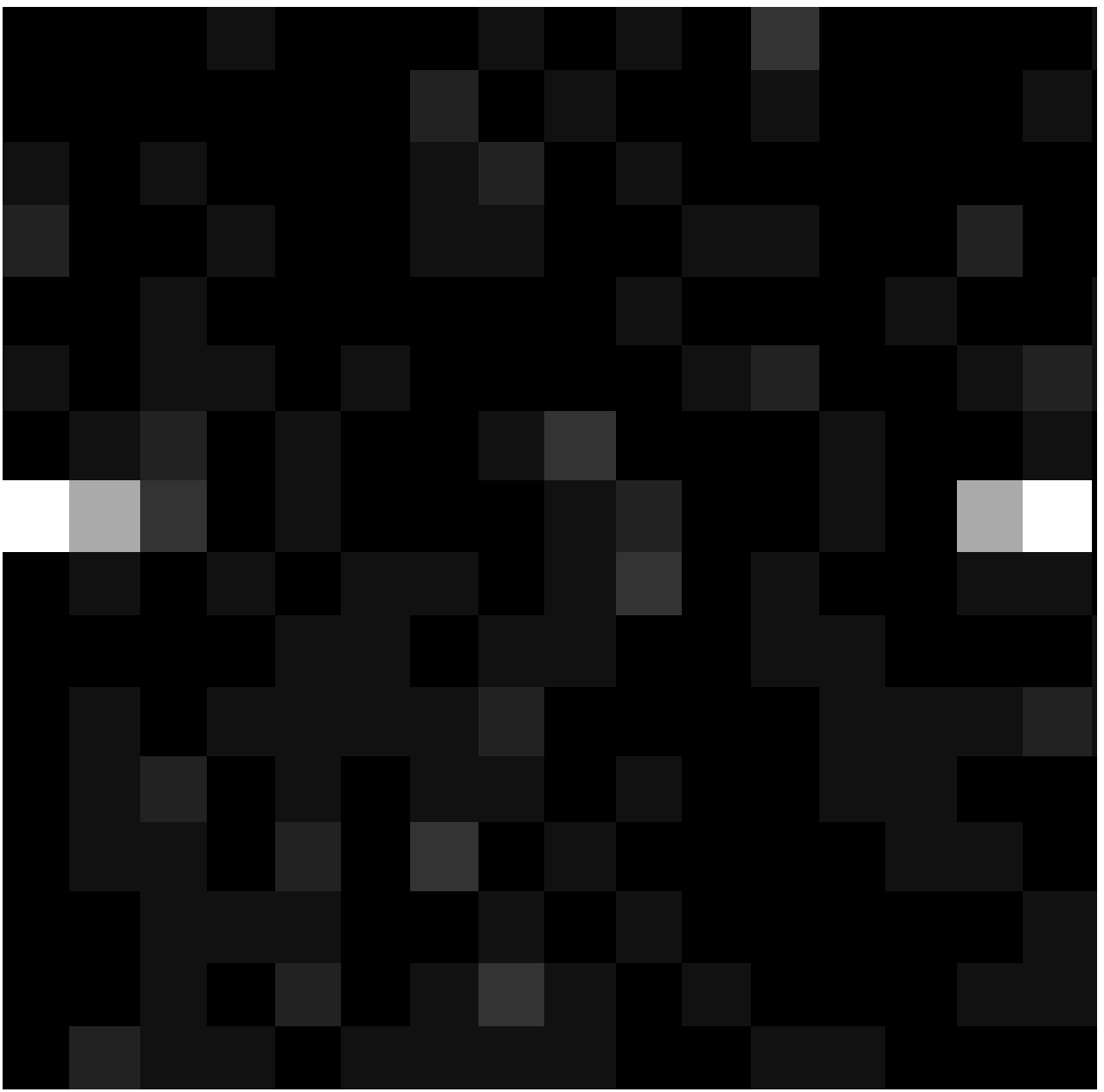}
\vspace{0.3 cm}
\includegraphics[height=3.0in,width=3.5in]{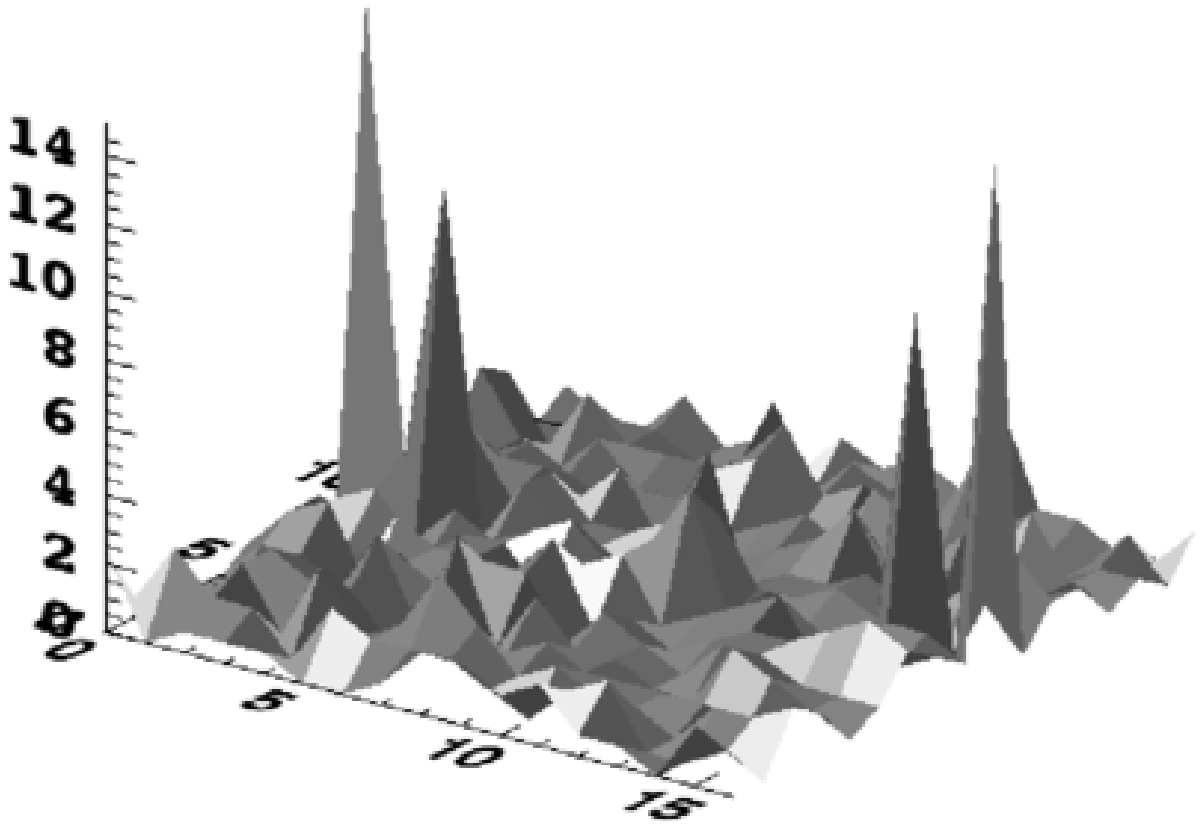}
\caption{(a) Fringes for two closely placed sources in the limited FOV of CONFIG-3
(top left). (b) Sources appear to be very close and just resolved in the reconstructed 2D (top 
right) and 3D (bottom) images. Sources at right side of central point in reconstructed plane 
represent the actual sources. Pseudo source (ghost image) is also seen in the reconstructed image plane 
along with the background noise.}
\end{figure}

\subsection{CONFIG-4: FZP2 + CMOS}

CONFIG-4 is the best possible configuration for imaging of a hard X-ray source in terms of achievable 
angular resolution. This configuration consists of dual FZP (FZP2) coder with a finest zone width of
0.0041 cm and high position sensitive CMOS detector with smallest pixel size of 0.005 cm.   
The FZP coder is of negative cosine type. Inner zone radius of each zone is 0.1 cm and number of 
zones is 144.

\begin{figure}[h]
\centering
\includegraphics[height=2.1in,width=2.1in]{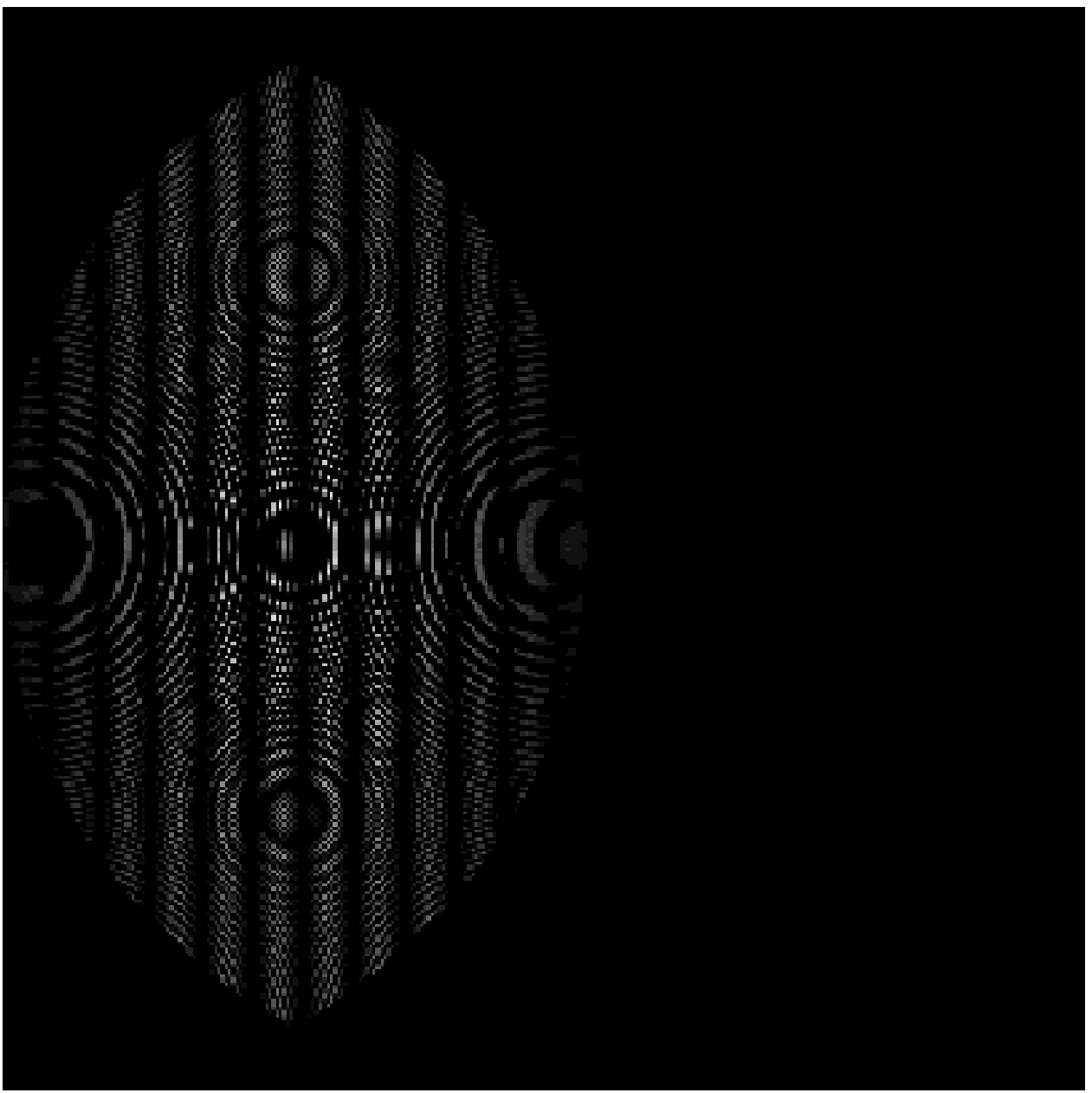}\hspace{0.1 cm}
\includegraphics[height=2.1in,width=2.1in]{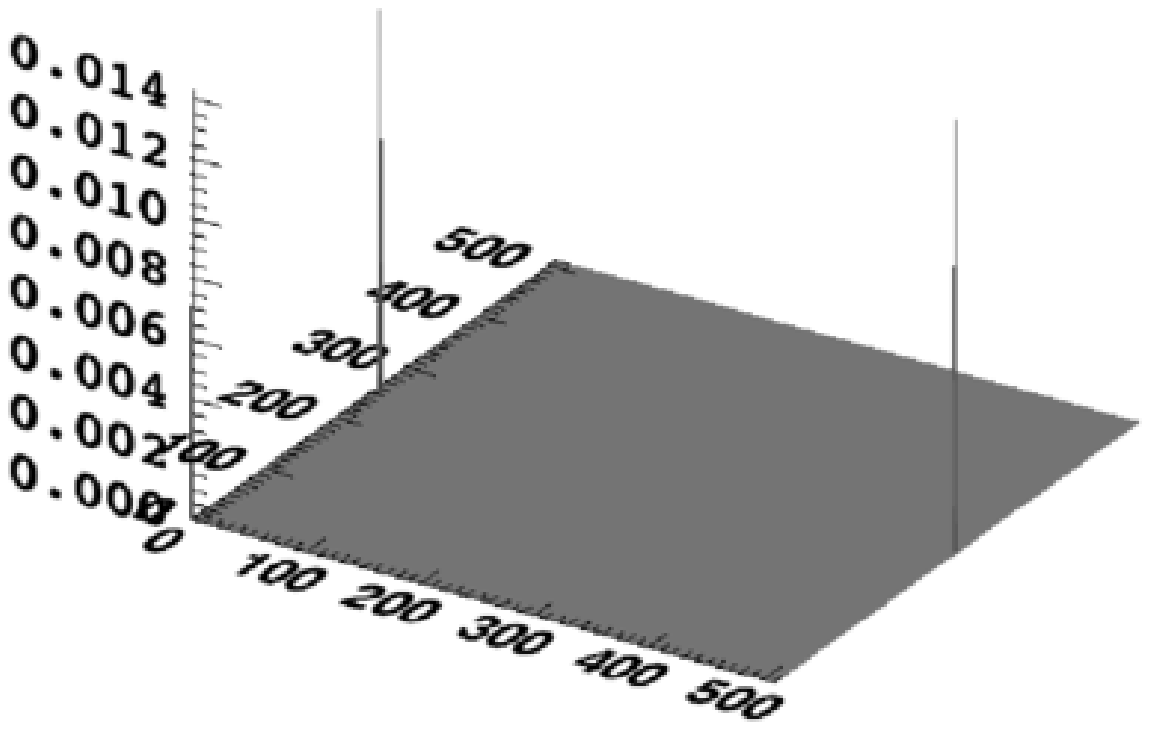}
\caption{(a) Fringes obtained on a CMOS detector with a pair of zone plates as coded aperture 
(left). (b) 3D picture of the reconstructed source along with pseudo source which also appears in 
the reconstructed source plane. The central DC offset is chopped out.}
\end{figure}

\begin{figure}[h]
\centering
\includegraphics[height=2.0in,width=2.0in]{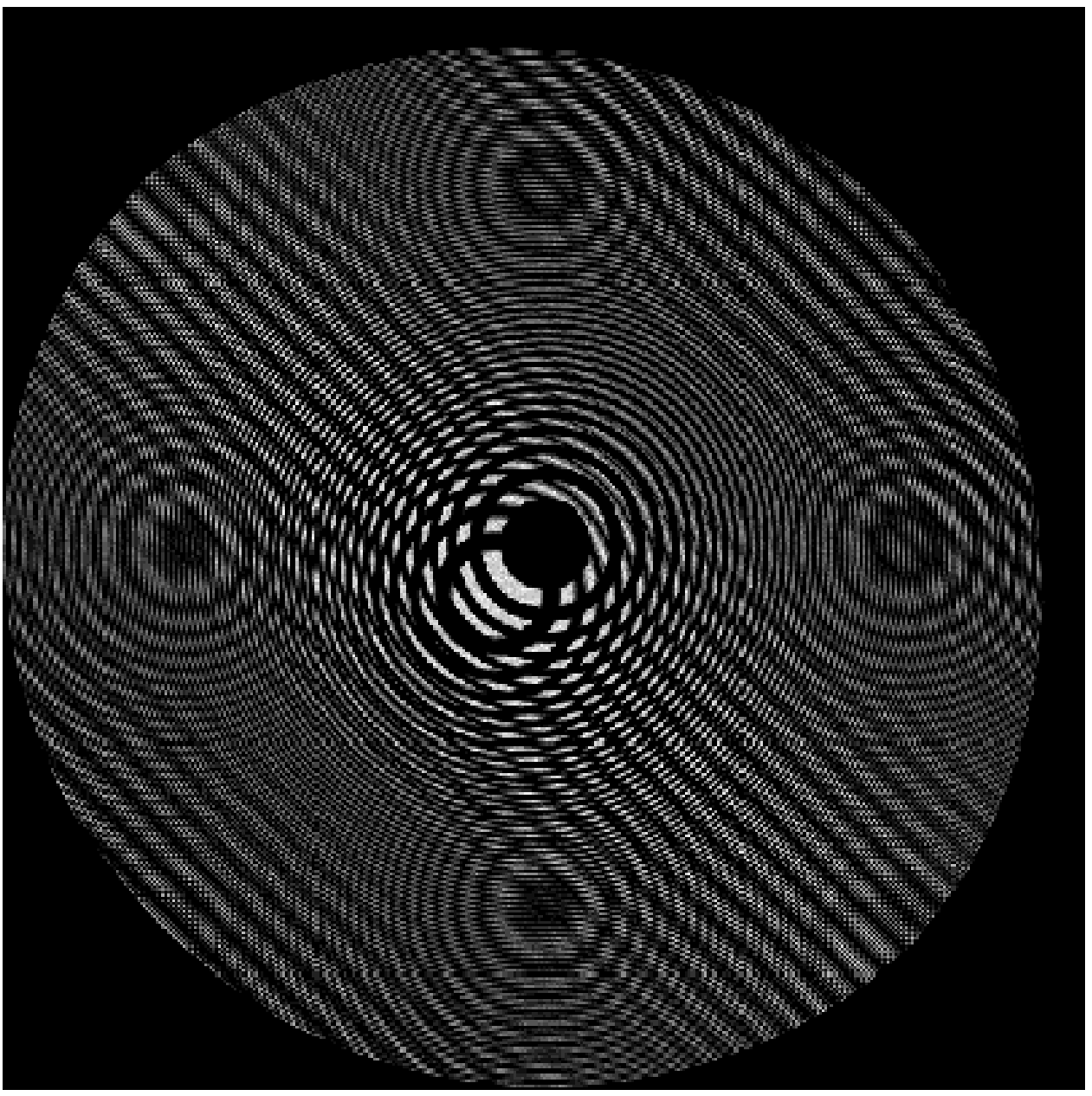}\hspace{0.1 cm}
\includegraphics[height=2.6in,width=2.6in]{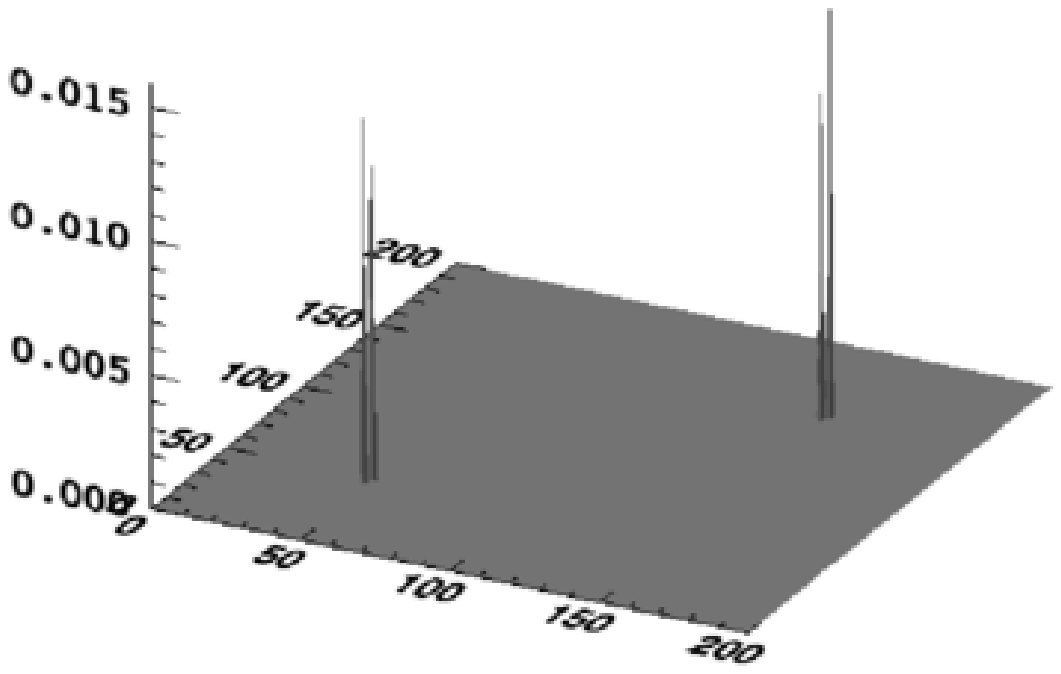}
\caption{(a) Fringes obtained (left) with a pair of sources at angular distance from each other 
equal to the calculated angular resolution ($54''$) of the FZP2-CMOS combination. (b) 3D view of 
the reconstructed sources. To get a closer view, the part of the 
reconstructed plane containing the sources is zoomed, so that out of 600 hundred pixels along each 
side only 200 pixels are shown.}
\end{figure}

\begin{figure}[h]
\centering
\includegraphics[height=1.7in,width=1.7in]{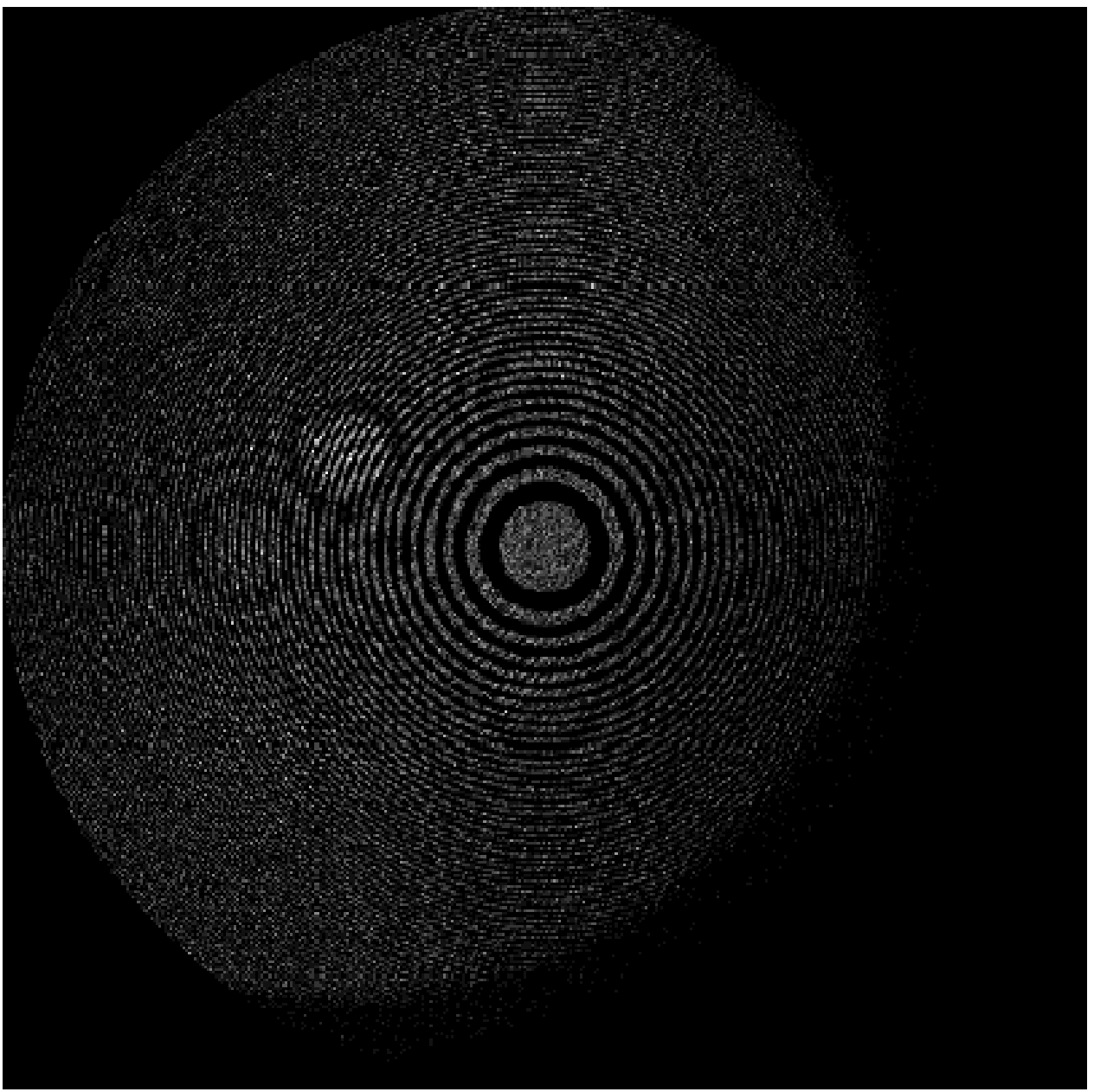}\hspace{0.1 cm}
\includegraphics[height=1.8in,width=1.8in]{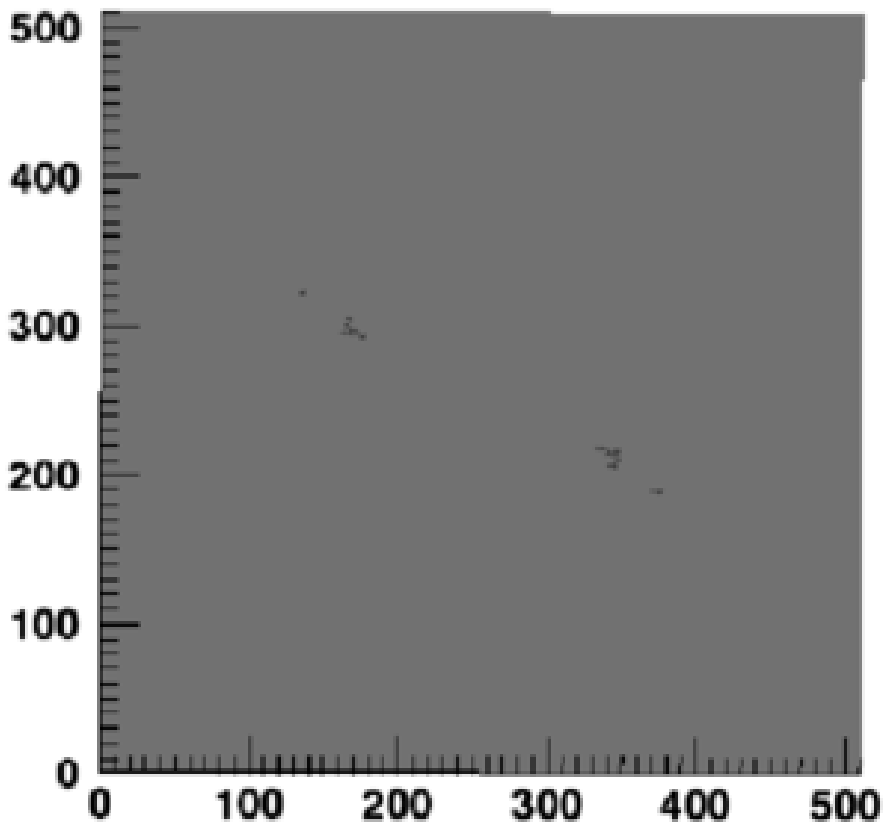}
\includegraphics[height=2.5in,width=3.0in]{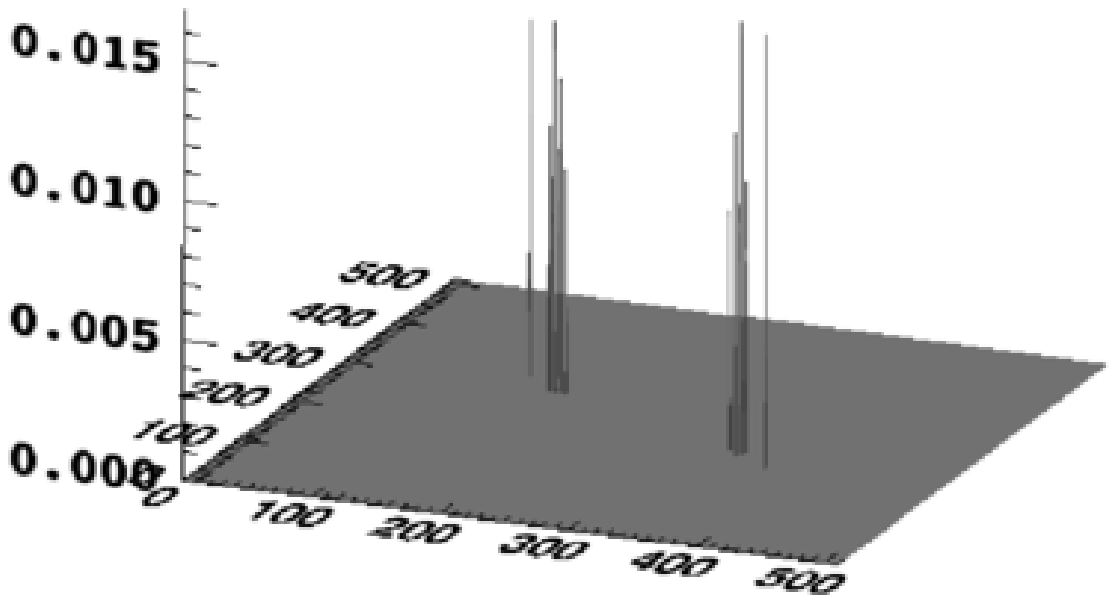}
\caption{(a) Fringes obtained with a pairs of zone plates on CMOS detector for multiple sources
(extended source) (top left). (b) 2D view of the reconstructed sources (top right). (c) 3D view of 
reconstructed sources of the sky plane.}
\end{figure}

The simulated Moir\'{e}  fringe pattern for on and off-axis source position is shown in Figure 1.
Now, we simulate for FOV verification of this configuration. We have considered a single source
with offset angle $\phi$ = 2.145$^\circ$ and $\theta$ = 0$^\circ$.

The  Moir\'{e} fringe pattern for the source with offset $\phi$ = 2.145$^\circ$ is shown in Figure 11(a). 
In Figure 11(b), 3D view of the reconstructed source plane is shown along with the pseudo source
(ghost source). Reconstructed source plane shows that the source is at extreme end of the FOV 
of the collimator, which confirms that the FOV is actually 4.29$^\circ$ (twice the $\phi$ value).

The most important aspect of this configuration is the best possible angular resolution which could be
around $54''$. To verify the mathematically calculated angular resolution value, we
simulate with two sources placed $54''$ apart and less than that. From simulation, it is found that
the sources which are placed less than $54''$ apart are not resolvable at all. In Figure 12(a,b),
we have shown the fringe pattern and reconstructed sky plane of two sources which are separated
by $54''$. The double pseudo source (ghost image) is also seen in the 3D view. Separation 
between the two closely placed sources are found to be equal to one detector pixel dimension.  

So far, we have considered a point source to do the simulation. In principle, it is also possible
to do simulation for extended sources. Simulation is done for the same configuration (CONFIG-4) with 
large number of point sources, which can be conveniently taken as an extended source. 
The fringe pattern along with 2D and 3D views of reconstructed sources are given in Figures 13(a-c).
Amplitudes of individual reconstructed source give their relative intensities in the extended source 
distribution.

\section{Expectations in the context of solar observations}

The prime objective of RT-2 Experiment onboard CORONAS-PHOTON satellite is to study the solar hard 
X-ray emission associated with solar flares. 
Solar flares are the most powerful explosions on the Sun, when the stored
energy in twisted magnetic field is suddenly released. Solar flares are generally
classified according to their X-ray brightness in the wavelength range of
1 to 8 {\AA}. Intensity of the classified flares (eg. A, B, C, M and X class) are measured
based on the peak flux (in unit of $W/m^{2}$) which is measured on the GOES
satellite. X-class flares are the most intense having peak flux of $10^{-4}$
$W/m^{2}$, while A-class flares are weakest of having peak flux of $10^{-8}$ $W/m^{2}$.
The peak flux of each class is 10 times greater than the preceding one with
a linear division of 9 within each class. Therefore, a C4.0 class flare is 11
times more powerful than a B3.0 class flare.

The spectral and temporal characteristics of solar hard X-ray flare are diverse, ranging from 
relatively soft, thermal (kT $\approx$ 10 keV) spectra, to hard, 
power law spectra, and from strong micro-flares with duration of seconds to events lasting 30 minutes 
or more. Therefore, the evolution of hard X-ray emission regions with time is an important aspect 
for better understanding of physics involved in the emission process. To pin-point the emission
regions, it is essential to have high angular resolution instruments. The RT-2/CZT payload which is
designed with four different configurations (in terms of AR and FOV) will serve the main purpose
to image the solar flares with high angular resolution.

%The main purpose of this experiment is to study the characteristics of solar hard X-ray 
%bursts with high time resolution spectral studies. 

Normally, the Sun is quiet to the level of being invisible in the hard X-ray and $\gamma$-ray energy 
band. The hard X-ray and $\gamma$-ray detection of Sun is necessarily a study of energetic solar 
flares of flare strength not less than B3.0. 

%The highest energy flares, expected to be seen  once every 50 days would
%give 2 counts in 50 msec above the average background of about 1 count.
%This will allow the study of time evolved spectral studies of several millisecond duration and search 
%for structures in solar hard X-ray bursts. 
%A continuous observation of one year should be able to 
%record about 10,000 bursts during a solar maximum, and the frequency is expected to be a factor 
%of 10 lower for solar minimum. This experiment is expected to generate a large database for 
%statistical analysis of the solar bursts along with detailed individual time structures and spectral 
%studies for about 100 solar flares. 
%As far as imaging is concerned this mission is first of it's 
%kind as it makes uses of CAM and FZP coder together as shadow caster first time in space based imaging.

We have considered that the maximum number of photons hitting the coder surface is around 
5$\times 10^5$ (at least for one source), while performing the simulations for all four 
configurations. In a realistic situation, like a solar flare of X4.8 type, which occurred on 
2002 July 23, (during the previous solar cycle),
we have estimated the response of the particular flare for all four configurations. The spectrum of
the giant flare during impulsive phase is double power-law like nature (Lin et al. 2003) with indices 
$\gamma_{L}\approx 6.2$ below and $\gamma_{H}\approx~ 2.8$ above a sharp break around 30 keV with a
maximum flux around 5~$\times 10^3 photons/sec/cm^2/keV$ at 20 keV. Using the above spectral 
information, we have calculated the number of photons impinging on each coders (four configurations) 
for 100 sec accumulation (onboard accumulation time of each image frame) at different energy bands. 
Apart from the real observation, we have also calculated the number of photons impinging on each 
coder surface for a typical flare of C3.5 type, which is characterized by a spectrum of single power 
law of index $\gamma \approx 4.0$ and flux (maximum) around 200~$photons/sec/cm^2/keV$ at 20 keV.
The estimated photon counts which impinges on all four configurations for both the flares, are 
given in the Table 4.

%In the table, we give estimations of number of photons 
%falling on each coder ($N_{ij}$) per 100 seconds (accumulation time for imaging in RT-2) in certain 
%energy ranges($E_{i}$ to $E_{j}$) for the X4.8 type flare which occurred on July 23, 2002. The 
%Here, the terminal energies for the calculation of approximate no of photons within them, 
%those are impinging on the coders within certain time span, are taken to be $E_{1}$= 20 keV for 
%lower and $E_{2}$ = 50 keV and $E_{3}$ = 100 keV for the upper limits. 

%\begin{center}
%\begin{tabular}{ l|c|c|c }
%\hline
%Configurations  & $ Counts $ & $ Counts$ &  Counts   \\
%	        & $ (20 - 50 keV) $ & $ (50 - 100 keV) $ & (20 - 100 keV) \\
%\hline
%\hline
%{\bf CONFIG-1}  & 1.22 $\times$ $10^7$ & & 1.43 $\times$ $10^7$  \\           % 12180800  14286400
%{\bf CONFIG-2}  & 1.22 $\times$ $10^7$ & & 1.43 $\times$ $10^7$ \\        % 12180800  14286400
%{\bf CONFIG-3}  & 5.38 $\times$ $10^6$ & & 6.31 $\times$ $10^6$  \\           % 5378584  6308338
%{\bf CONFIG-4}  & 3.44 $\times$ $10^6$ & & 4.04 $\times$ $10^6$  \\           % 3442294 4037336   
%\hline
%\end{tabular}
%\end{center}

\begin{center}
\centerline {Table 4}
\begin{tabular}{ l|c|c|c|c|c }
\hline
{\bf Class} & {\bf Energy} & {\bf CONFIG-1} & {\bf CONFIG-2} & {\bf CONFIG-3} & {\bf CONFIG-4}   \\
\hline
      & (keV)  &    Counts      &   Counts       &   Counts       &     Counts        \\
\hline     
      & (20-50)  & 1.22 $\times$ $10^7$ & 1.22 $\times$ $10^7$ & 5.38 $\times$ $10^6$ &  --- \\        
X4.8  & (50-100) & 2.12 $\times$ $10^6$ & 2.12 $\times$ $10^6$ & 7.31 $\times$ $10^5$ & --- \\    
      & (20-100) & 1.43 $\times$ $10^7$ & 1.43 $\times$ $10^7$ & 6.31 $\times$ $10^6$ & 4.04 $\times$ $10^6$ \\    
\hline
      & (20-50)  & 6.24 $\times$ $10^5$ & 6.24 $\times$ $10^5$ & 2.75 $\times$ $10^5$ &  --- \\        
C3.5  & (50-100) & 3.70 $\times$ $10^4$ & 3.70 $\times$ $10^4$ & 1.62 $\times$ $10^4$ & --- \\    
      & (20-100) & 6.61 $\times$ $10^5$ & 6.61 $\times$ $10^5$ & 2.91 $\times$ $10^5$ & 1.88 $\times$ $10^5$ \\    
\hline

\end{tabular}
\end{center}

It is evident from the estimations given in Table 4, that we could detect clear and prominent fringes for 
flares like X4.8 and as low as C3.5 type flare with all four configurations. 
%also for less intense flares (eg. B type flare) should be obtained with the different configurations 
%that are employed in RT-2/CZT payload. 
For less intense flares as low as B type flares, though there may not be images 
with prominent fringes in the detectors but after proper reconstruction it could still be possible to
reproduce the sources in the field of view.

The FOV (5.72$^\circ$) of CAM and CZT detector (CONFIG-1 and CONFIG-2) is enough to
accommodate the whole Sun. Due to poor angular resolution ($21.5'$), it is hardly 
possible to observe and distinguish more than one flare, simultaneously.
Imaging of a single flare with CONFIG-3 is more critical as the 
FOV in this configuration is very small ($409''$). Therefore, a highly pointed observation 
(within $409''$)
of flare with this configuration is possible to image with moderate angular resolution ($64''$).
On the other hand, the imaging of full Sun with hard X-ray solar flares is nicely possible with
the CONFIG-4, as the FOV (4.57$^\circ$) and angular resolution ($54''$) are superior compared to 
any other configuration.

\section{Laboratory test results}

Tests with CONFIG-2 (CONFIG-1) set up were carried out at laboratory of VSSC, 
Thiruvananthapuram, India. The set up consists of single plane coder (CAM2) and CZT (CZT2) detector. 
The CAM is shined with a strong radio-active source placed on top of the collimator at position of 
$\theta$=150$^\circ$ (approx) and $\phi$=1.04$^\circ$. In Figure 14a (top left), 
the shadow pattern of the CAM obtained in the CZT detector is shown. 
%In Figures 14(a-c) we have the result of experiment done with the CAM presented in Figure 6(a)
%Figure 14a(top left) is the shadow pattern obtained in the CMOS detector. 
In Figures 14b (top right) and 14(c, d) (bottom-left: source with background noise, bottom-right: 
background noise is averaged out), we have shown the 2D and 3D view of the reconstructed 
source plane. The source in the reconstructed plane (detector plane) is nearly exact reproduction 
(in terms of position and intensity) of the original source. The source, though 
actually a point source, is spread over two pixels of the reconstructed image plane. 
It is due to the divergent
nature of rays from the radio-active source, which impinges on the coder (CAM2), placed at 132 cm
away from the detector plane.  

\begin{figure}[h]
\centering
\includegraphics[height=1.8in,width=1.8in]{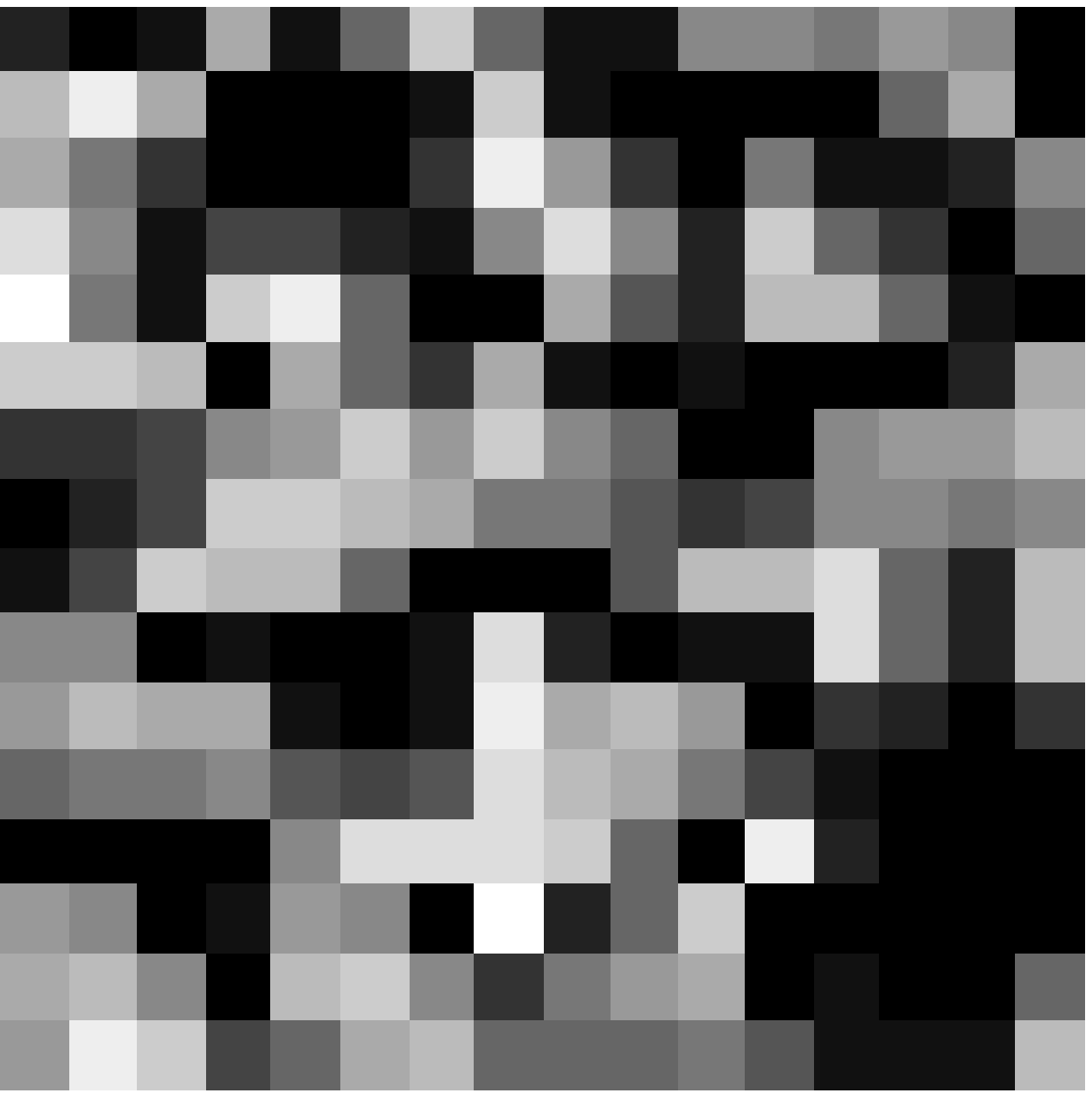}  \hspace{0.2 cm}  
\includegraphics[height=1.8in,width=1.8in]{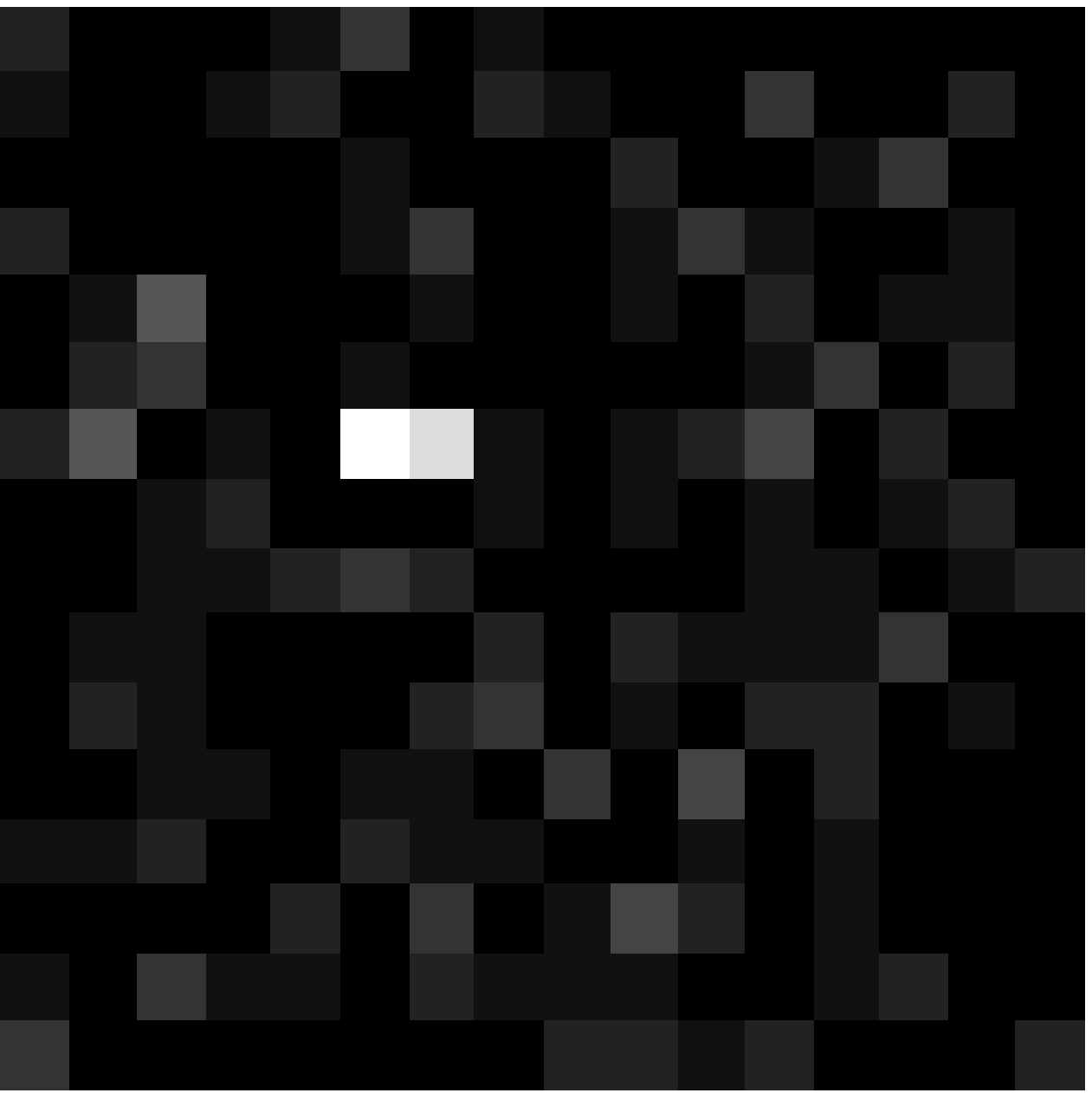}
\includegraphics[height=2.0in,width=2.3in]{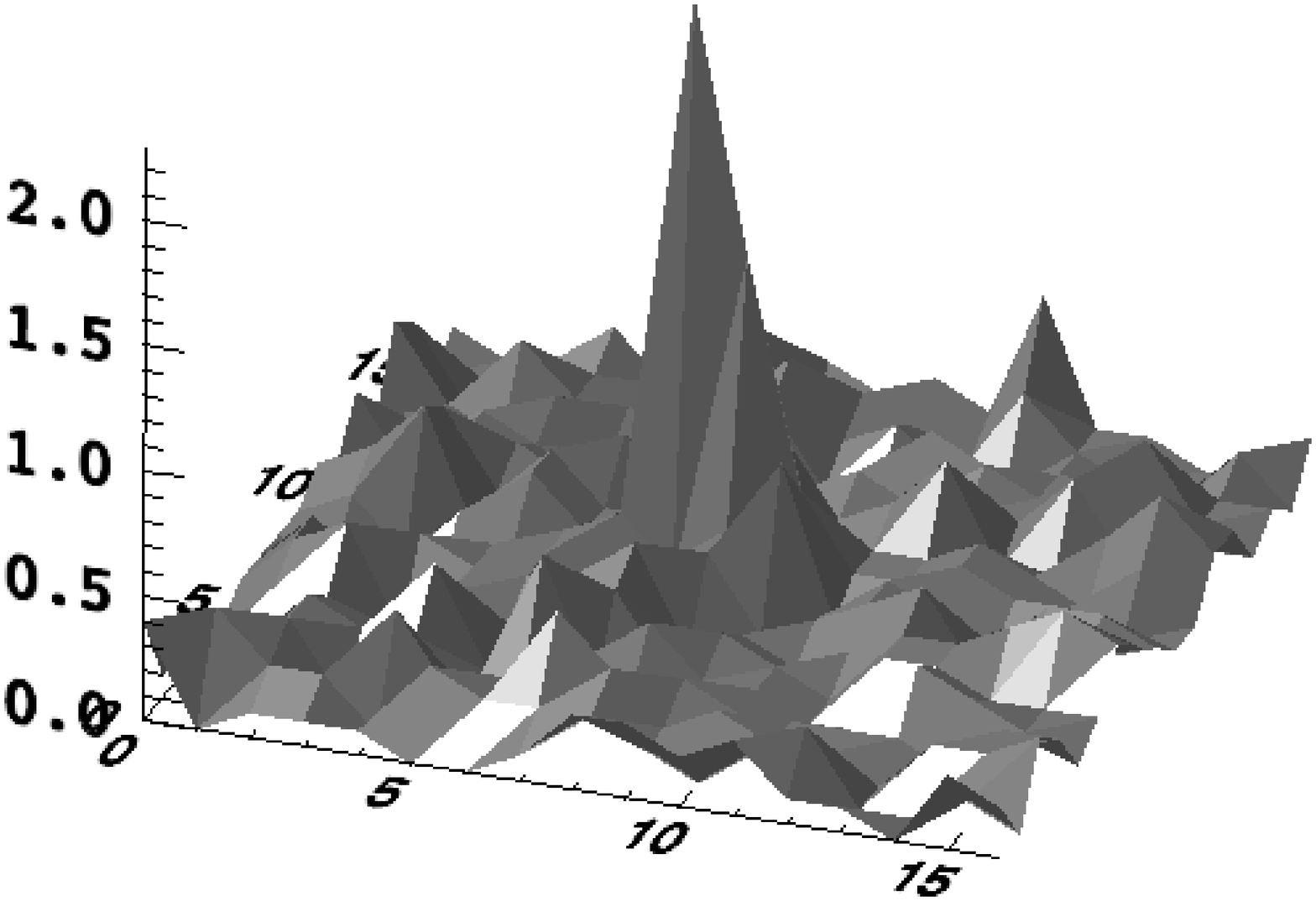}
\includegraphics[height=2.0in,width=2.3in]{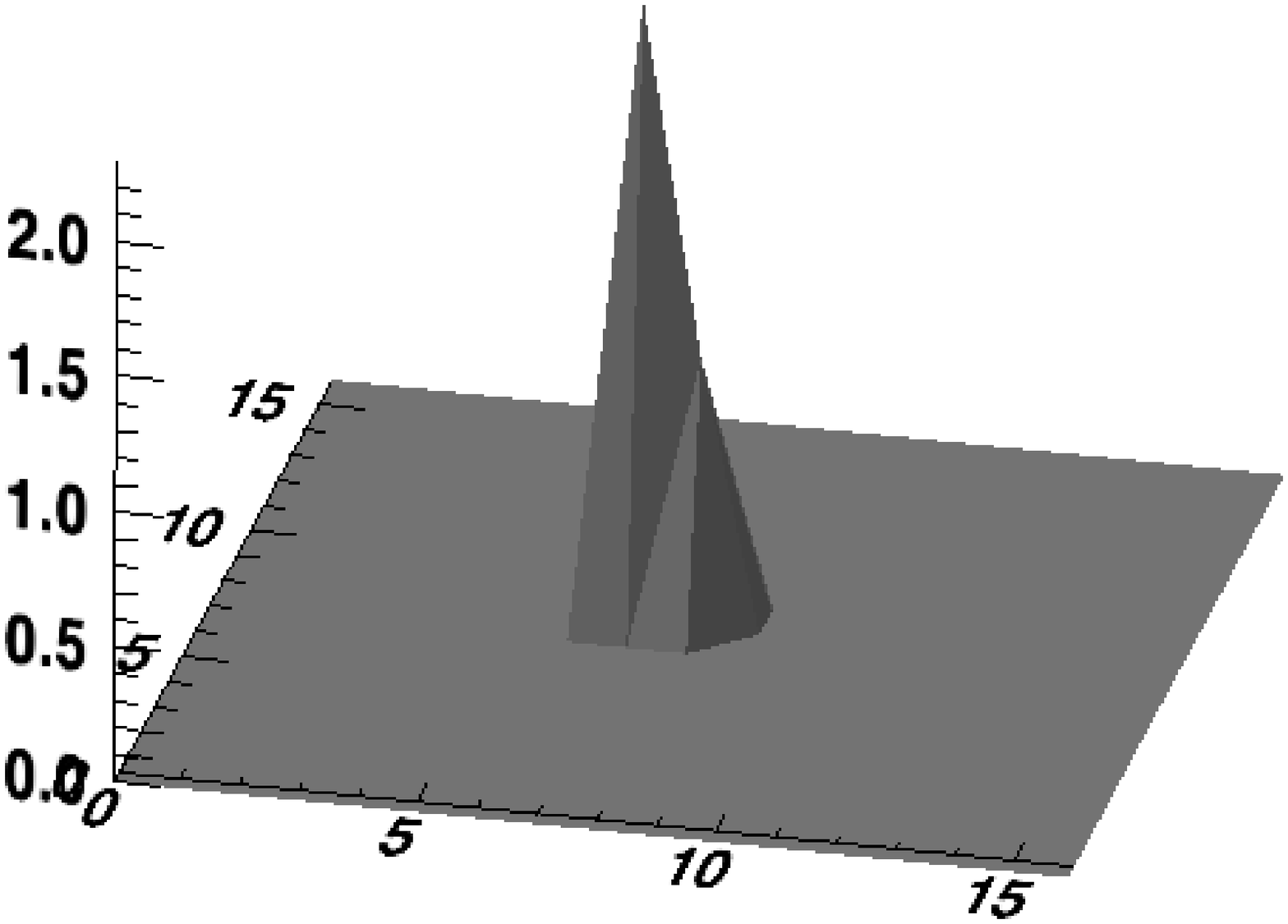}
\caption{(a) Shadow pattern obtained during tests with a CAM on a CZT detector for a single 
source (top-left). (b) 2D picture of the reconstructed source (top-right). (c) 3D picture of reconstructed 
source with background noise. (d) 3D view of the reconstructed source after smoothing the background
noise.}
\end{figure}

Tests  with FZP set up were carried out at the X-ray laboratory of ICSP, Kolkata, India which is 
equipped with an X-ray source generator of operating voltage 5 Volt to 50 Volt. As it is difficult to 
produce parallel X-ray beam at laboratory, we have generated quasi-parallel (diverging) X-ray beam
with 45 feet collimator made of lead shielded aluminum pipe. The detector system (collimator having
FZPs and detector) is kept at one end of the 45 feet long pipe opposite to the X-ray source.

%Below some experimental results with FZP coders are presented. Figure 15(a) gives the fringes 

%obtained for a positive cosine Fresnel zone plate coder. 

In the first set of experiment, we have taken dual zone plates of negative in 
nature and the n-th zone radius for each is $\sqrt{n}$ times the inner zone radius. The inner zone 
radius is 0.1 cm and number of zones in each zone plate is 144 with finest zone width of 0.0041 cm. 
The separation between the zone plates is 32 cm. Highly position sensitive CMOS detector is used with 
smallest pixels size of 0.005 cm. 

\begin{figure}[h]
\centering
\includegraphics[height=2.0in,width=2.0in]{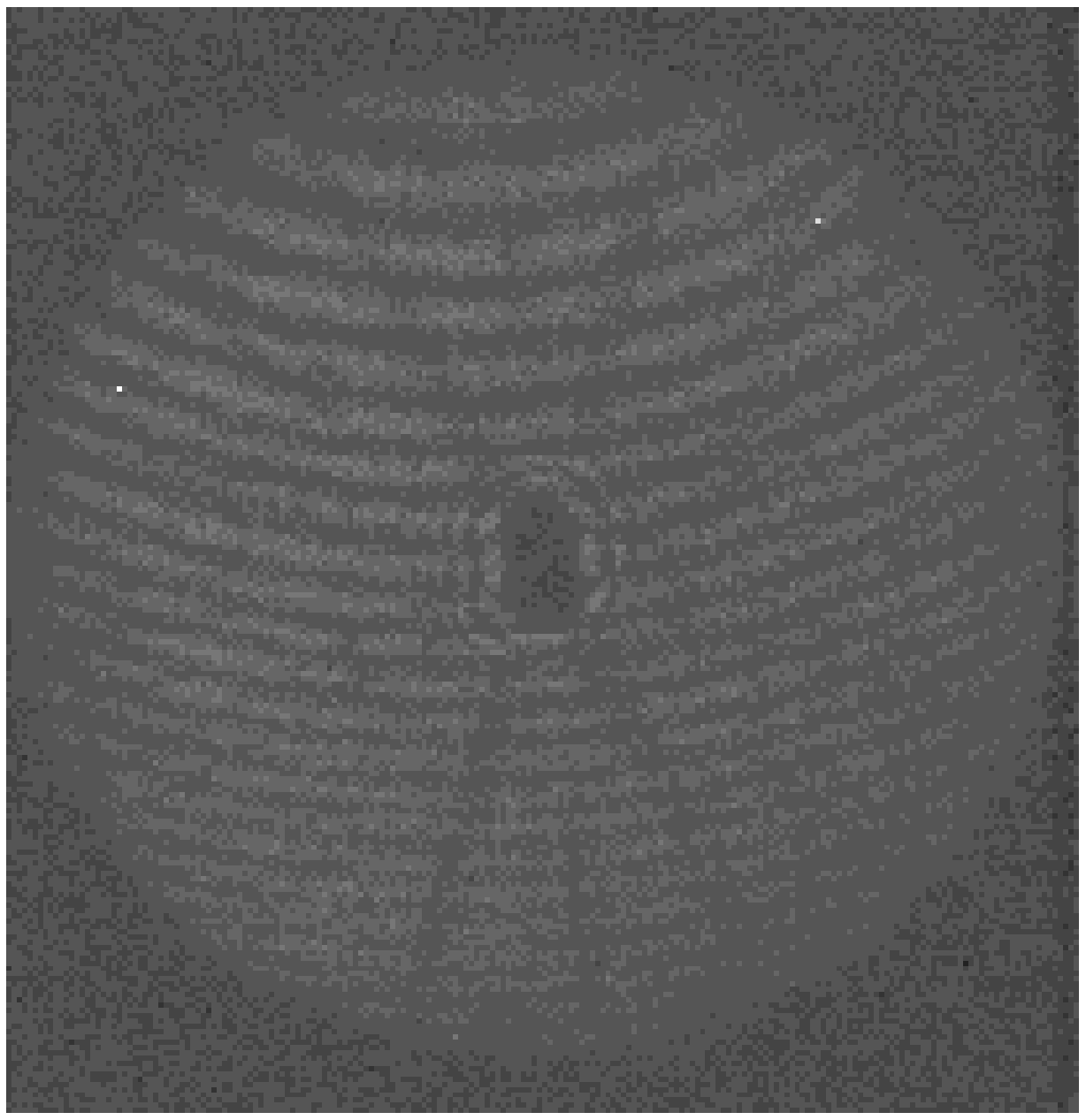}\hspace{0.2 cm}     % 11
\includegraphics[height=2.2in,width=2.2in]{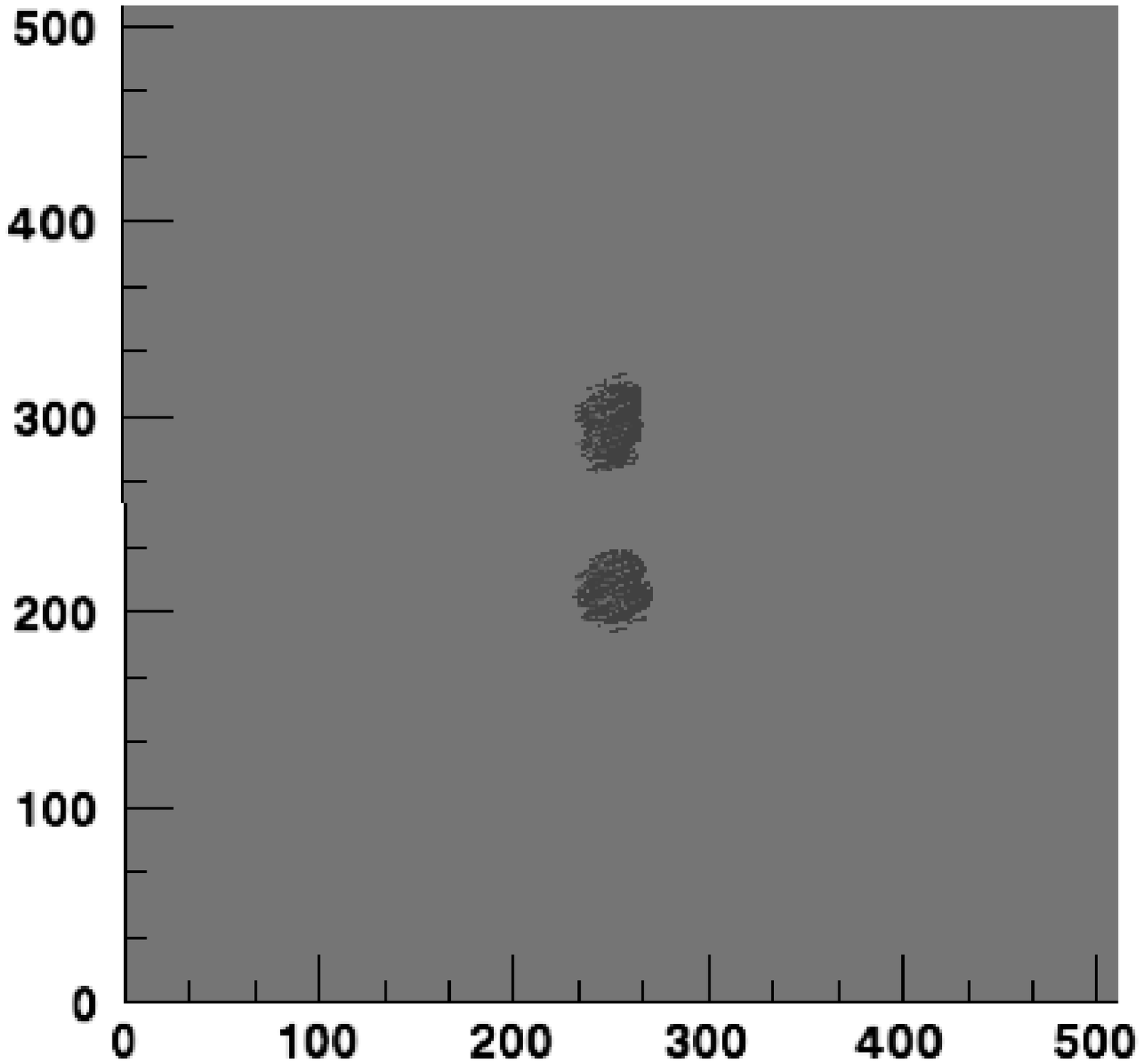}
\includegraphics[height=2.4in,width=2.4in]{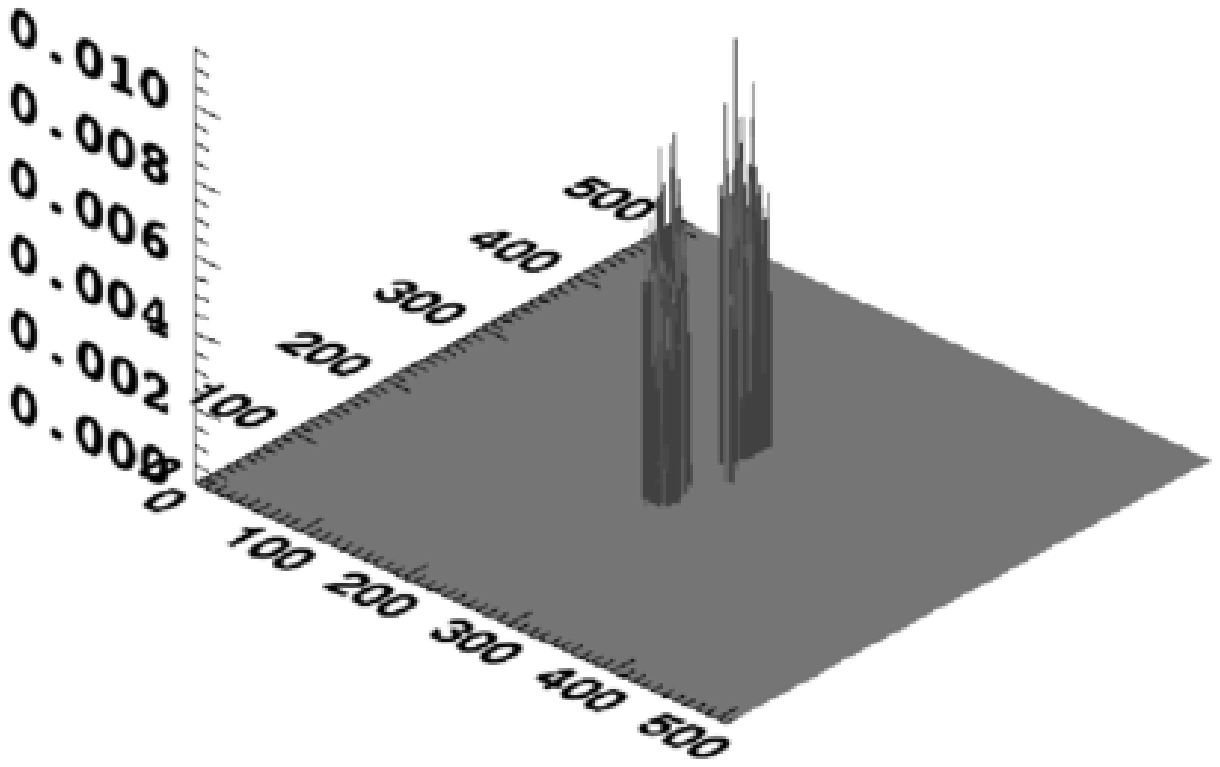}
\caption{(a) Fringes obtained with a pair of negative cosine FZP on CMOS detector for a 
slightly off-axis source (top left). (b) 2D view of the reconstructed source (top right) and (c) 3D 
view of reconstructed source are seen in image plane along with pseudo source (bottom).}
\end{figure}

The X-ray source is positioned at a little off-axis with the central line of beam. The Moir\'{e} fringe 
pattern observed in the detector plane (CMOS) is shown in Figure 15(a). In Figure 15(b,c), we have 
plotted the 2D and 3D view of the reconstructed point source in the detector plane. 
The pseudo source (ghost image) and central DC offset both appear in the experimental results. 
As the central DC-offset is incomparably bigger than the reconstructed source and its pseudo part, 
we have to chop it out to get the source prominently. Measurement gives the off-axis
angle of the 
source ($\phi$) as $32'$ $20''$ and $\theta$ as 93$^\circ$. This is consistent with the priorly specified 
actual source position. The divergence of the projected X-ray photon beam has caused broadening of 
reconstructed source and makes it look like a circular spot.
 
In another experimental set up, different coder (FZP1) along with CMOS detector is used. The coder is 
cosine type and positive in nature.
%Another experimental result is given in Figure 16(a-c). Here another set of zone plates is used as 
The inner zone radius of each of the zone plate is 0.122 cm and number of zones is 151 with finest
zone width 0.0050 cm. The observed Moir\'{e} fringe pattern on the detector plane is shown in Figure 
16(a). The reconstructed 2D and 3D images in detector plane is shown in Figure 16(b,c). Actual
source (right side) along with the pseudo source (ghost image) is seen in the image.
The source is offset by an amount $\phi$=$21'$ $30''$ and at $\theta$=50$^\circ$ $31'$. 
The reconstructed 2D (top right) and 3D pictures (bottom left) represent exact replica of the source 
plane.    

\begin{figure}[h]
\centering
\includegraphics[height=4.0in,width=2.0in]{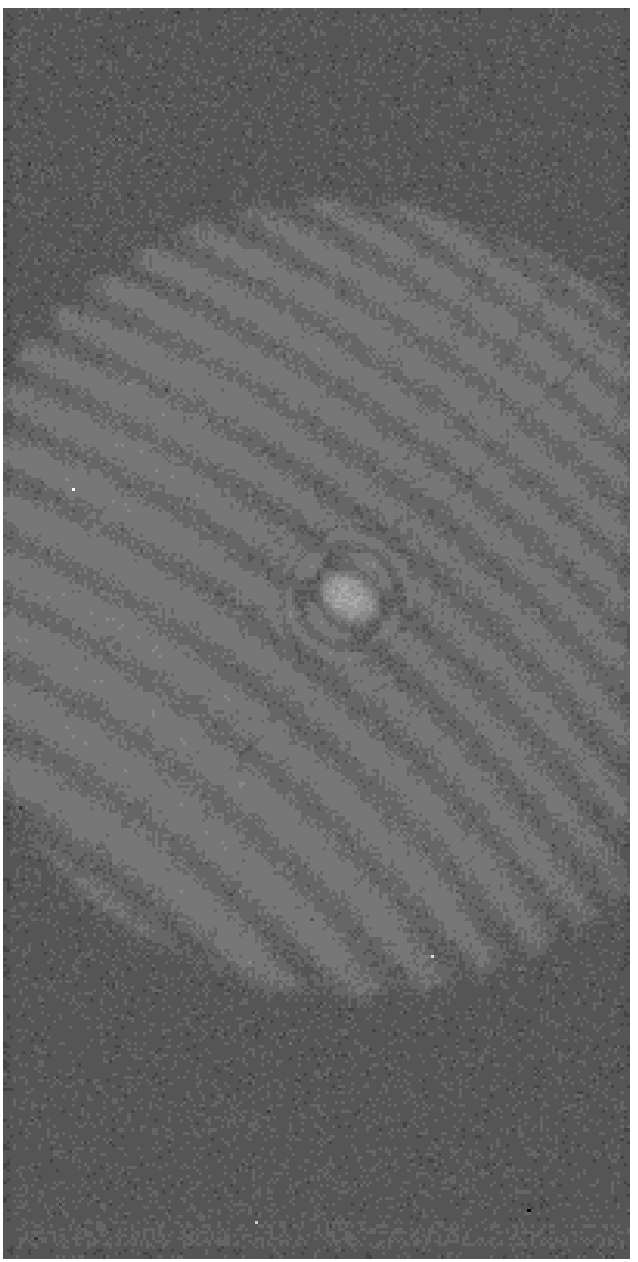}\hspace{0.2 cm}       %12
\includegraphics[height=3.7in,width=2.5in]{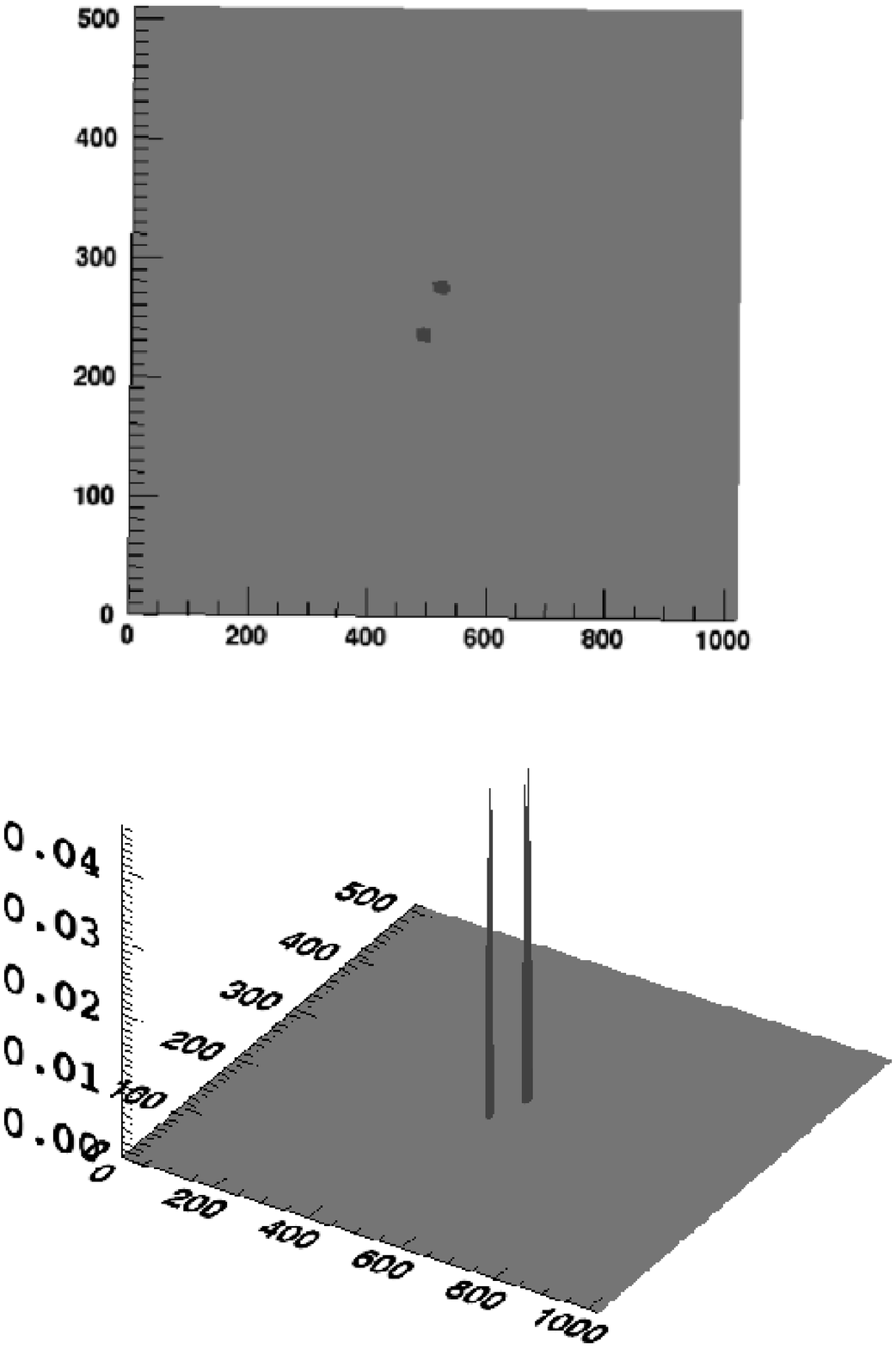}

\caption{(a) Fringe obtained experimentally with a pair of zone plates (FZP1) on CMOS detector 
for a source (top left) at offset $21'$ $30''$. (b) 2D view of the reconstructed source (top right) and 
(c) 3D view of reconstructed source are seen along with pseudo source.}
\end{figure}

\section{Conclusion}

The RT-2/CZT payload onboard CORONAS-PHOTON mission is a unique instrument for imaging in hard X-rays.
It uses four different kinds of configurations with which various combinations of angular resolutions (AR)
and FOVs are achievable for imaging a single source at a time. This instrument also uses FZP coder as
a shadow casting device for imaging in hard X-rays in space flight, for the first time. Two different
types of coders (CAM and FZP) along with two different types of detectors (CZT and CMOS) are used to achieve
variable angular resolutions. As far as imaging is concerned, this mission is first of its 
kind as it makes use of CAM and FZP coder together as shadow caster for the first time in space based 
imaging.

The FZP coders, used in the CONFIG-4 along with high position sensitive CMOS detector (50 $\mu$), have 
angular resolution which is much better than those of the configurations consisting CAMs and CZTs. The
best geometric resolution of this instrument is around $54''$. 
The hard X-ray imaging devices rely on the number of photons detected, and
hence the centroiding accuracy for point sources can be much better than this,
and can reach upto a few arc-seconds for very bright sources. 
For example, the RHESSI satellite uses another alternative approach to image in hard X-rays with
Rotation Modulation Collimator (RMC) technique and, depending on the
source intensity, source localization accuracy ranged from $2''$ to $180''$. 
 
It is also possible to achieve good spatial resolution with CAM-CMOS configuration. For that
one has to design the CAM elements size comparable to those of CMOS pixels to maintain the consistency 
of spatial resolution with that of the detector. 
This is a much more complex effort in terms of fabrication and alignment.

In this paper, we have presented simulation results along with direct 
reconstruction of images to quantify the AR and FOV of the various 
configurations used in the payload. Some of the simulation results
are verified with laboratory measurements. Since all the 4 detectors would
be simultaneously imaging the same event (viz., solar flares), it 
should be possible to make a simultaneous fit to all the four
images with a few assumed source positions and hence to accurately 
measure the source positions and intensity. This will mitigate some
of the problems in the direct reconstruction like ghost images in the
FZP images and noise patterns in the CAM images. 

On 30th January, 2009, the CORONAS-PHOTON was launched successfully 
and all the RT-2 instruments are working to our satisfaction. However, so far, in the
24th cycle, the solar activity has been weak and we are awaiting stronger flares 
for direct imaging. The on-board data quality and results would be discussed elsewhere.

\begin{acknowledgements}
SP and DD thank CSIR/NET scholarships and RS and TBK thank
RT-2/SRF fellowship (ISRO) which supported their research work.
The authors are thankful to scientists, engineers and technical staffs from TIFR/ICSP/VSSC/ISRO-HQ for 
various supports during RT-2 related experiments.
\end{acknowledgements}

\end{document}